\documentclass[%
reprint,
superscriptaddress,
amsmath,
amssymb,
aps,
rmp,
floatfix,
]{revtex4-2}

\usepackage{graphicx}
\usepackage{dcolumn}
\usepackage{bm}
\usepackage{hyperref}
\usepackage[mathlines]{lineno}

\usepackage{stmaryrd} 
\usepackage{xcolor, soul}
\usepackage{placeins} 
\usepackage{subcaption} 
\usepackage{enumerate} 
\usepackage{yhmath} 

\begin{document}
\preprint{APS/123-QED}

\title{Revisiting Elastic String Models of Forward Interest Rates}

\author{Victor Le Coz}
\email{victor.lecoz@gmail.com}
\affiliation{Quant AI lab, 29 Rue de Choiseul 75002 Paris, France}
\affiliation{Chair of Econophysics and Complex Systems, \'Ecole polytechnique, 91128 Palaiseau Cedex, France}
\affiliation{LadHyX UMR CNRS 7646, \'Ecole polytechnique, 91128 Palaiseau Cedex, France}

\author{Jean-Philippe Bouchaud}
\email{jean-philippe.bouchaud@cfm.com}
\affiliation{Chair of Econophysics and Complex Systems, \'Ecole polytechnique, 91128 Palaiseau Cedex, France}
\affiliation{Capital Fund Management, 23 Rue de l’Universit\'e, 75007 Paris, France}
\affiliation{Académie des Sciences, Quai de Conti, 75006 Paris, France}

\date{\today}

\begin{abstract}
Twenty five years ago, several authors proposed to describe the forward interest rate curve (FRC) as an elastic string along which idiosyncratic shocks propagate, accounting for the peculiar structure of the return correlation across different maturities. In this paper, we revisit the specific ``stiff'' elastic string field theory of \citet{BaaquieBouchaud-2004} in a way that makes its micro-foundation more transparent. Our model can be interpreted as capturing the effect of market forces that set the rates of nearby tenors in a self-referential fashion. The model is parsimonious and accurately reproduces the whole correlation structure of the FRC over the time period~$1994-2023$, with an error around $1\%$ and with only one adjustable parameter, the value of which being very stable across the last three decades. The dependence of correlation on time resolution (also called the Epps effect) is also faithfully reproduced within the model and leads to a cross-tenor information propagation time on the order of $30$ minutes. Finally, we confirm that the perceived time in interest rate markets is a strongly sub-linear function of real time, as surmised by \citet{BaaquieBouchaud-2004}. In fact, our results are fully compatible with hyperbolic discounting, in line with the recent behavioral Finance literature \citep{FarmerGeanakoplos-2009}.
\end{abstract}


\maketitle



\section{Introduction} \label{Introduction}

\subsection{Motivation}

The forward interest rate $f(t,T)$, to be defined more precisely below, is the interest rate agreed upon at time $t$, for an instantaneous loan between $T \geq t$ and $T+dT$. Such a collection of future rates defines a kind of ``string'' that moves and deforms with time. Understanding the dynamics of the forward interest rate curve (FRC) is crucial in a wide spectrum of financial applications, ranging from the valuation of interest rate derivatives to risk management \citep{Hull-2018, BrigoMercurio-2006}. This problem is also fascinating from a theoretical point of view: whereas the stochastic process governing the dynamics of single assets (point-like objects) has been thoroughly investigated (see e.g. \citep{Bachelier-1900,Osborne-1959,BlackScholes-1973,Heston-1993,BacryEtAl-2001,GatheralEtAl-2018,Zumbach-2010,DandapaniEtAl-2021,WuEtAl-2022,morelEtAl-2023}), the stochastic process of higher dimensional objects like lines or graphs is much more involved \citep{AiharaBagchi-2005,CarmonaTehranchi-2006, EkelandTaflin-2005, Filipovic-2001}. There is a long tradition in the physics literature of modeling string-like (or surface-like) objects which has not yet pervaded into the financial mathematics literature, despite early attempts \cite{BouchaudEtAl-1999, Santa-ClaraSornette-2001, BaaquieBouchaud-2004}. 

The aim of this paper is to revisit the 2004 proposal of Belal Baaquie and one of the author (JPB), to describe the returns of different tenors of the FRC in terms of the fluctuations of a ``stiff'' elastic string -- called henceforth the BB04 model \cite{BaaquieBouchaud-2004}. We will see that up to a redefinition of their model that accounts for the discrete set of maturities defining the FRC (instead of the continuum limit of BB04), the proposed framework allows one to account quite remarkably for the full cross-maturity correlation structure of the FRC, across the whole period 1994-2023 and a single adjustable parameter -- when the BB04 model was only tested for the period 1994-1996 and had three parameters.

\subsection{Definitions and notations}

We recall here the definition of the risk-free instantaneous forward interest rate. Table~\ref{tab:notations} in appendix~\ref{Notations} provides the complete list of the notations used in this study.

\paragraph{Zero-coupon Bond.}
Let~$P(t,T)$ represent the price at time~$t$ of a zero-coupon bond maturing at~$T$. Such a bond pays one unit of currency at maturity~$T$ without any intermediate coupons.

\paragraph{Forward Rate.}
Consider time~$t$ and two future times, $S$ and~$T$, where $t<S<T$. The forward rate, a risk-free interest rate for the period~$[S,T]$, is derived from zero-coupon bonds. By selling a zero-coupon bond maturing at $S$ for $P(t,S)$ euros and purchasing~$\frac{P(t,S)}{P(t,T)}$ units of a bond maturing at $T$, we establish a contract that costs nothing at $t$, pays one unit of currency at $S$, and yields $\frac{P(t,S)}{P(t,T)}$ euros at $T$. This setup leads to a deterministic rate of return, with the continuously compounded forward rate~$R(t,S,T)$ given by:
\begin{align}
e^{R(t,S,T)(T-S)} = \frac{P(t,S)}{P(t,T)},
\end{align}
solving to:
\begin{align}
R(t,S,T) = -\frac{\log{P(t,T)} - \log{P(t,S)}}{T-S}.
\end{align}

\paragraph{Instantaneous Forward Rate.}
As $S$ approaches $T$, the limit of $R(t,S,T)$ defines the instantaneous forward rate~$f(t,T)$:
\begin{align}
    f(t,T) = -\frac{\partial \log{P(t,T)}}{\partial T},
\end{align}
The collection of these rates for various~$T$ forms the forward rate curve (FRC).

In the following sections, we actually define the instantaneous forward rate~$f(t,\theta)$ in terms of the time to maturity or \textit{tenor}~$\theta = T-t$. This dimension~$\theta$ is often referred to as the \textit{space} dimension, as opposed to the time dimension~$t$.

\subsection{The Heath-Jarrow-Morton framework}

The Heath-Jarrow-Morton (HJM) framework has become the industry standard \citep{HeathEtAl-1992,Hughston-1996}. Within this framework, the FRC dynamics is described by Itô processes driven by a $d$-dimensional Brownian motion. Consequently, bond prices for each tenor~$\theta$ are regarded not as financial derivatives of the risk-free rate~$f(t,0)$ but as individual risky assets, leading to an possibly infinite number of such assets. The finite number~$d$ of diffusion factors introduces the potential for arbitrage opportunities among bond prices \citep{Bjork-2019}. Thus, conditions are established on the drift components of instantaneous forward rate processes to ensure arbitrage-free pricing of zero-coupon bonds. This framework rests only on two fundamental assumptions: the continuity of sample paths for forward rate processes and a finite number of Brownian motions driving these processes.

Beyond the fact that the HJM model has no ambition to capture the ``physical'', one dimensional nature of the FRC, a limitation of the HJM framework is its stipulation that for any integer~$k$, the correlation matrix of~$k+d$ instantaneous forward rates must be singular when the model employs~$d$ factors \citep{Goldstein-2000}. This condition is in conflict with empirical observations. Addressing this limitation, various researchers have ventured beyond the conventional boundary of a finite number of driving Brownian motions. Notably, \citet{Kennedy-1994,Kennedy-1997} proposed to simulate each forward rate by a Gaussian random field while \citet{Cont-2005a}, \citet{Goldstein-2000} and \citet{Santa-ClaraSornette-2001} developed stochastic string approaches, partly based on the empirical work of \cite{BouchaudEtAl-1999} where the idea of the FRC as an elastic string was first put forth.

Among these advancements, \citet{Baaquie-2001,Baaquie-2002,Baaquie-2004} has pioneered a  field theory approach, set to be further discussed in the ensuing section. In the following years, these random field theories have been applied to solve interest rate derivatives pricing problems \citep{Bueno-GuerreroEtAl-2015,Bueno-GuerreroEtAl-2016,Bueno-GuerreroEtAl-2020,Bueno-GuerreroEtAl-2022, Baaquie-2007,Baaquie-2009, Baaquie-2010, Baaquie-2018, BaaquieTang-2012, BaaquieLiang-2007, WuXu-2014}.

\section{A field theory for the FRC}

\citet{Baaquie-2001,Baaquie-2002,Baaquie-2004} introduced a two-dimensional field theory to describe the forward interest rate curve. This approach was generalized in BB04 \citet{BaaquieBouchaud-2004} to account for the pronounced smoothness observed in the correlation matrix of forward rate increments. More precisely, it was observed in \cite{BouchaudEtAl-1999} that the eigenvectors of the covariance matrix of the FRC returns had the same structure as those of a elastic string and can be indexed by the number $k$ of zeroes in the $\theta$ direction. The corresponding eigenvalues were found to behave as $(a + bk^2)^{-1}$ for small $k$ (where $a,b$ are constants), crossing over to a faster decay $\approx k^{-4}$ for larger $k$ (see Fig. 5 of \cite{BouchaudEtAl-1999}).

A way to encode this empirical finding is to posit that the dynamics of the FRC $f(t,\theta)$ is specified by a drift velocity~$\gamma(t,\theta)$ and a volatility~$\sigma(t,\theta)$, such that \citep{BaaquieBouchaud-2004}:
\begin{align}
    \frac{\partial f}{\partial t} (t,\theta) &= \gamma(t,\theta) + \sigma(t,\theta) A(t,\theta),
\end{align}
where $A(t,\theta)$ represents a driftless (Langevin) noise field. 

The ``field theory'' formulation assumes that $\theta$ is a continuous variable $\in \mathbb{R}_{+}$ and the joint probability distribution across time and tenor for the set of noises $\left\{A(t,\theta)\right\}_{(t,\theta)\in \mathbb{R}_{+}^2}$ is determined by the exponential of an action $S[A]$. This action is a functional defined over the semi-infinite domain $\mathbb{R}_{+}^2$, and is given by \citep{BaaquieBouchaud-2004}:\footnote{Note that throughout this paper, we only speak of {\it classical} (i.e. non quantum) field theories, describing the statistical physics of extended objects.}
\begin{align}
    \label{eq:action}  \nonumber
    S[A] := &-\frac{1}{2} \iint_0^\infty {\rm d}t {\rm d}\theta \, \Bigg(  \Bigg. A^2(t,\theta) 
    \\   
    & +  \left( \frac{1}{\mu} \frac{\partial A}{\partial \theta} (t,\theta)\right)^2 +  \left(\frac{1}{\nu^2} \frac{\partial^2 A}{\partial \theta^2}(t,\theta)\right)^2 \Bigg. \Bigg), 
\end{align}
with~$\mu^{-2}$ and~$\nu^{-4}$ denoting respectively the ``line tension'' and ``stiffness'' (also called bending rigidity) parameters, which have the physical dimensions of frequencies. More precisely, small values of $\mu$ disfavor large local slopes of $A(t,\theta)$ whereas small values of $\nu$ disfavor large local curvatures. A boundary condition is needed for the theory to be complete, and was postulated in \cite{BaaquieBouchaud-2004} to be of the Neumann type, i.e.
\begin{align}
    \label{eq:bc}
    \left. \frac{\partial A (t,\theta)}{\partial \theta}\right|_{\theta=0} &= 0,
\end{align}
thereby enforcing a uniform motion of the forward interest rates at very short maturities. This assumption is justified considering the spot rate $f(t, 0)$ is typically set by the Central Bank, and very short-term maturities carry minimal additional risk.

Note that the absence in Eq. \eqref{eq:action} of any coupling between different infinitesimal {\it time} slices (i.e. along the $t$ direction) means that $A(t,\theta)$ has no temporal correlations and behaves as a white noise. More precisely, ${\rm d}W(t,\theta) := A(t,\theta) {\rm d}t$ is a standard Wiener noise, with non trivial covariance along the $\theta$-direction 
\begin{align} \label{eq:Adelta}
    \langle {\rm d}W(t,\theta) {\rm d}W(t',\theta') \rangle &= \mathcal{D}_{BB}(\theta,\theta') {\rm d}t,
\end{align}
where the brackets denote an averaging over the functional weight $e^{S[A]}$ and where  $\mathcal{D}_{BB}(\theta,\theta')$ is found to be \citep{BaaquieBouchaud-2004}
\begin{align} \label{eq:BB04}
    \mathcal{D}_{BB}(\theta,\theta') &= \frac{\nu^4}{\alpha_+-\alpha_-}\left[\frac{F(\theta,\theta',\sqrt{\alpha_-})}{\alpha_-} - \frac{F(\theta,\theta',\sqrt{\alpha_+})}{\alpha_+}\right],
\end{align}
with 
\[ 
\alpha_{\pm} = \frac{\nu^4}{2 \mu^2}\left[ 1 \pm \sqrt{1-4\left(\frac{\mu}{\nu}\right)^4}\right]
\] 
and
\begin{align}
    F(\theta,\theta',p) &:= \frac{p}{2}\left( e^{-p(\theta+\theta')}+e^{-p|\theta-\theta'|}\right).
\end{align}

Formally, absence of arbitrage among zero-coupon bond prices imposes the following condition on the drift $\gamma(t,\theta)$ \citep{Baaquie-2004}:
\begin{align}
    \gamma(t,\theta) &= \sigma(t,\theta) \int_{0}^{\theta} {\rm d}\theta' \mathcal{D}_{BB}(\theta')\sigma(t,\theta'),
\end{align}
but this term is usually completely negligible numerically \citep{BouchaudEtAl-1999}, and we will drop it henceforth.

Baaquie and Bouchaud further introduced the concept of \textit{psychological} time which explains how the perceived time $\theta'-\theta$ between tenors varies with their distance $\theta$ from the observer standing at time $t$, introducing one more parameter called ${\psi}$ below (see section~\ref{Psychological_time} below). This framework allows one to fit the {\it whole} empirical correlation matrix $\rho_{\theta \theta'}$ with only three meaningful parameters $\mu, \nu$ and ${\psi}$. Within this model, and in line with observations, the {\it curvature} of the forward rate correlation perpendicular to the diagonal, decays as power-law with respect to maturity (see Appendix~\ref{Curvature along the anti-diagonal}). This was perhaps the most salient success of the BB04 model, which however fell into almost complete oblivion (only 12 citations to date!).

\begin{figure}
    \centering
    \includegraphics[width=\linewidth]{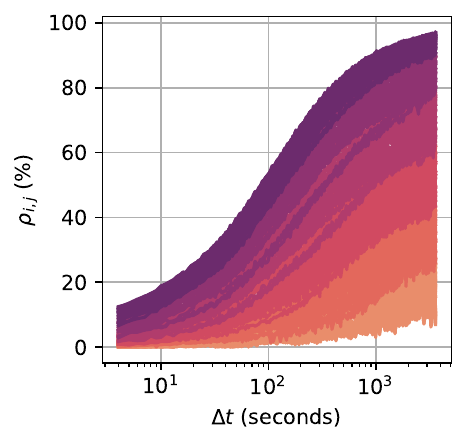}
    \caption{Pearson correlation coefficients~$\rho(\theta,\theta')$ as a function of the time scale~$\Delta t$ used to define returns, for each pair of SOFR Futures prices of time-to-maturity $\theta$ ranging from $3$-month to $60$-month for the year $2021$. Each color corresponds to a pair ordered from the lowest (orange) to the highest (purple) correlation level at the $1$-hour bin.}
    \label{fig:df_epps_corr}
\end{figure}

In spite of its phenomenological success, the BB04 model has two main limitations. 
\begin{enumerate}
    \item First, the theory assumes a continuous spectrum of futures contracts across different tenors $\theta$, whereas in reality, futures contracts are available only at discrete tenors, usually every three calendar months. In other words, continuous derivatives like $\partial_\theta A$ have no physical existence. 
    \item Second, it predicts a constant correlation structure across all time scales used to define returns. This contradicts the well known ``Epps effect'', i.e. the influence of the temporal granularity used to analyze prices \citep{Epps-1979,Reno-2003,TothKertesz-2009}. Specifically, for the SOFR futures prices, Fig.~\ref{fig:df_epps_corr} illustrates that the correlation between pairs of tenors ranging from 3 to 60 months emerges only after several minutes.
\end{enumerate} 

The aim of the present paper is to revisit the BB04 model with the two above deficiencies in mind. We reformulate the model in a way that makes its micro-foundations more apparent. We show in particular that market participants enforce a self-referential dynamic to the FRC, where each tenor is directly influenced by the motion of its neighbors. Furthermore, our dynamical formulation encapsulates market micro-structure phenomena, including the Epps effect, non-martingale prices at short scales and price-impact and cross-impact effects. The latter will be detailed in a companion paper \citet{LeCozEtAl-2024}.

\FloatBarrier
\section{A Dynamical Reformulation}

\subsection{The continuous limit}

We now want to interpret the above driftless noise field~$A(t,\theta)$, as the solution to a stochastic Langevin equation that will generate temporal correlations over a time scale~$\tau \ll 1$ day, instead of instantaneous correlations as in Eq. \eqref{eq:Adelta}.

We then define~$\eta(t,\theta)$ to be a two-dimensional Gaussian (Langevin) noise, characterized by the covariance function
\[
\langle \eta(t, \theta) \eta(t',\theta') \rangle = 2 D \delta(t-t') \delta(\theta-\theta'),
\]
where $\delta(\cdot)$ represents the Dirac delta function and $2D$ is the variance of the noise. [In order to give a precise meaning to the following expressions, we adopt the physicists' convention and consider ${\rm d}t$ and ${\rm d}\theta$ to be very small compared to all relevant time scales of the problem, but not infinitesimal, so that time and maturities can be thought as discrete but very densely probed.]  

We now postulate the following stochastic evolution for $A(t,\theta)$: 
\begin{equation}
    \label{eq:principle}
    \frac{\partial A}{\partial t}(t,\theta) = \frac{1}{\tau} \left( \frac{\delta S[A]}{\delta A(t,\theta)} + \eta(t, \theta)\right),
\end{equation}
where $\tau$ is the characteristic time scale for the emergence of correlations (see below). We will not attempt to define the ``functional derivative'' ${\delta S[A]}/\delta A(t,\theta)$ in a mathematically rigorous manner, but note that it boils down to the usual derivative if we think of time and maturities as discrete. 

Eq. \eqref{eq:principle}, alongside the Neumann boundary condition specified in Eq.~\eqref{eq:bc}, can be expressed through the following linear differential equation:
\begin{equation}
    \label{eq:master}
    \left\lbrace
    \begin{aligned}
    &\frac{\partial A}{\partial t}(t,\theta) = \frac{1}{\tau} \left[ -\mathcal{L}[A](t,\theta) + \eta(t, \theta) \right], \\
    &\frac{\partial A}{\partial \theta}(t,0) = 0, \quad \forall t
    \end{aligned}
    \right.
\end{equation}
where 
\[ 
\mathcal{L}[A](t,\theta):= A(t,\theta) - \frac{1}{\mu^2} \frac{\partial^2 A}{\partial \theta^2}(t,\theta) + \frac{1}{\nu^4} \frac{\partial^4 A}{\partial \theta^4}(t,\theta).
\]
Eq.~\eqref{eq:master} describes how the correlated noise field $A(t,\theta)$ responds to the uncorrelated shocks $\eta(t,\theta)$, for example the order flow at the microstructure level. An important property of the dynamics described by Eq. \eqref{eq:master} is that it leads, for a fixed single time $t \to \infty$, to a stationary measure $P[\{A\}]$ for $A(t,\theta)$ given by 
\begin{align}
    \label{eq:measure_action}  \nonumber
    P[\{A(t,\theta)\}] =&\exp \left[-\frac{1}{4D} \int_0^\infty {\rm d}\theta \, \Bigg(  \Bigg. A^2(t,\theta) 
    \right. \\   
    & \left. +  \left( \frac{1}{\mu} \frac{\partial A}{\partial \theta} (t,\theta)\right)^2 +  \left(\frac{1}{\nu^2} \frac{\partial^2 A}{\partial \theta^2}(t,\theta)\right)^2 \Bigg. \Bigg) \right], 
\end{align}
(see e.g. \citet{Kampen-2007} for a detailed discussion). In other words, for $2D=1$, the marginal of $e^{S[A]}$ for a given time $t$ with $S[A]$ defined in Eq. \eqref{eq:measure_action} coincides with $P[\{A\}]$. Therefore, all equal-time correlations of the field $A(t,\theta)$ coincide with the BB04 model. However, as we will discuss in section \ref{sec:discrete_d}, non-trivial temporal correlations develop when $\tau > 0$, and only disappear in the limit $\tau \to 0$ , which is the limit where BB04 is recovered.

\subsection{Psychological time}
\label{Psychological_time}

\citep{BaaquieBouchaud-2004} observed that the curvature of the forward rate correlations along the diagonal decays as a power law of the maturity (see Appendix~\ref{Curvature along the anti-diagonal}). To capture this behavior, BB04 proposed the change of variable~$\bar{z}(\theta) = \theta^{\bar{\psi}}$ with $\bar{\psi} < 1$ (see also \citet{BaaquieSrikant-2004}). This new variable, referred to as the \textit{psychological time}, ensures that the perceived time between events is a decreasing function of the maturity since ${\rm d}\bar{z} = \bar{\psi} \theta^{\bar{\psi}-1} {\rm d}\theta$. In other words, a month in a year appears longer than a month in ten years. 

In spite of its phenomenological success, this formulation violates the constraint that for very small maturities, psychological time and real time should become equivalent, i.e. tomorrow and a day after tomorrow are perceived (nearly) exactly the same way, whereas the above specification leads to a diverging value of ${\rm d}\bar{z}$ in that limit. Moreover, several studies in Neuroscience, Behavioral Economics or Finance \citep{GreenEtAl-1994,Sozou-1998,FrederickEtAl-2002,GreenMyerson-2004,Thaler-2005,DasguptaMaskin-2005,FarmerGeanakoplos-2009,KimZauberman-2009,RayBossaerts-2011,Cui-2011} suggest economic agents use hyperbolic discounting, which is tantamount to a {\it logarithmic} increase of the perceived time:\footnote{Such logarithmic form was also recently discussed by C. Tebaldi (unpublished). Note that if we insist on a regularized power-law dependence, $\bar{z}(\theta) = \bar{\psi}/\zeta \left(1 + \theta/\bar{\psi}\right)^{\zeta} - 1$, such that $\bar{z} \approx \theta$ when $\theta \to 0$, empirical calibration always returns a small value of $\zeta$, i.e. a logarithmic behaviour.} 
\begin{equation}
    z(\theta) = \psi \log\left(1 + \frac{\theta}{\psi}\right),
\end{equation}
which is such that $z(\theta) \approx \theta$ for $\theta \ll \psi$ and ${\rm d}{z} \approx \psi {\rm d}\theta/\theta$ for $\theta \gg \psi$. Indeed, an exponential discount with rate $r$ in psychological time reads
\begin{equation}
    e^{-rz(\theta)} = {\left(1 + \frac{\theta}{\psi}\right)^{-r \psi}},
\end{equation}
which coincides with hyperbolic discounting in real time.

Applying the  change of variable $z \to \theta(z) = \psi(e^{z/\psi} - 1)$ to the linear operator $\mathcal{L}[A](z,t)$ in Eq.\eqref{eq:master} yields a non-linear differential equation,
\begin{equation}
    \label{eq:master_general}
    \left\lbrace
    \begin{aligned}
    &\frac{\partial A}{\partial t}(t,\theta) = \frac{1}{\tau} \left[ -\mathcal{O}[A](t,\theta) + \eta(t, \theta) \right], \\
    &\frac{\partial A}{\partial \theta}(t,0) = 0, \quad \forall t,
    \end{aligned}
    \right.
\end{equation}
where $\mathcal{O}$ is the operator defined by
\begin{align}
    \mathcal{O}[A](t,\theta)&:= A(t,\theta) + \left(\frac{1}{\psi^3\nu^4}- \frac{1}{\psi\mu^2}\right) \left(1+\frac{\theta}{\psi}\right) \frac{\partial A}{\partial \theta}(t,\theta) \nonumber \\
    &+ \left(\frac{7}{\psi^2\nu^4}- \frac{1}{\mu^2}\right) \left(1+\frac{\theta}{\psi}\right)^2 \frac{\partial^2 A}{\partial \theta^2}(t,\theta) \nonumber \\
    &+ \frac{6}{\psi\nu^4}\left(1+\frac{\theta}{\psi}\right)^3 \frac{\partial^3 A}{\partial \theta^3}(t,\theta) \nonumber \\ 
    &+ \frac{1}{\nu^4} \left(1+\frac{\theta}{\psi}\right)^4\frac{\partial^4 A}{\partial \theta^4}(t,\theta).
\end{align}

The above formulation presumes a continuous (or at least very dense) spectrum of tenors $\theta$, whereas, in practice, forward rates are observed at discrete maturities only. As will be shown below, this is not a trivial difference. In order to enhance the realism of our model, we now explicitly discretize the above equation with respect to $\theta$. 

\subsection{A discrete counterpart}
In the following sections, we denote by $x_{\theta}$ any variable defined on the discrete space of tenors. $x(t)$ denotes the vector of components $x_{\theta}(t)$. Notably, $f(t,\theta)$ becomes $f_{\theta}(t)$ and $A(t,\theta)$ becomes $A_{\theta}(t)$. $f(t)$ and $A(t)$ are respectively the vectors of the forward rate and the noise field at time $t$.  We also define $I_k$ with $k \in \mathbb{Z}$ as a matrix of infinite size with ones ones only on the $k$-th diagonal above the main diagonal, i.e.,
\begin{equation}
(I_k)_{ij} =
    \begin{cases}
     1 \quad \text{if} \quad j-i = k, \\
     0 \quad \text{otherwise}.
    \end{cases}
\end{equation}
Note that $I_0$ is the identity matrix denoted simply $I$. Using a centered scheme, the discretization of Eq.~\eqref{eq:master_general} reads:
\begin{equation}
    \label{eq:discret_master_general}
    \left\lbrace
    \begin{aligned}
   & \frac{{\rm d}A}{{\rm d}t}(t)  = \frac{1}{\tau} \left[ -\mathcal{M} A(t)+\eta(t)\right] \\
    &A_{1}(t) - A_{-1}(t)= 0,
    \end{aligned}
    \right.
\end{equation}
where $\mathcal{M}$ is a matrix of infinite size defined by
\begin{align}
    \mathcal{M}_{\theta\theta'} &:= (I)_{\theta\theta'} \nonumber \\
    &+ \left(\frac{1}{\psi^3\nu^4}- \frac{1}{\psi\mu^2}\right) \left(1+\frac{\theta}{\psi}\right)\left(\frac{1}{2}I_1 - \frac{1}{2}I_{-1} \right)_{\theta\theta'} \nonumber \\
    &+ \left(\frac{7}{\psi^2\nu^4}- \frac{1}{\mu^2}\right) \left(1+\frac{\theta}{\psi}\right)^2 \Big( I_1 - 2I +I_{-1}\Big)_{\theta\theta'}  \nonumber \\
    &+ \frac{6}{\psi\nu^4}\left(1+\frac{\theta}{\psi}\right)^3 \left(\frac{1}{2}I_{2} - I_1 + I_{-1} - \frac{1}{2}I_{-2}\right)_{\theta\theta'}  \nonumber \\ 
&+ \frac{1}{\nu^4} \left(1+\frac{\theta}{\psi}\right)^4\Big( I_{2} - 4I_1 + 6I -4I_{-1} + I_{-2} \Big)_{\theta\theta'}.
\end{align}
In the above equation, $\theta \in \mathbb{N}$ is counted in multiple of 3 months and $\mu$ and $\nu$ are now dimension-less. The discretization of Eq.~\eqref{eq:master_general} also requires to replace the continuous noise~$\eta(t,\theta)$ by a discrete Langevin noise~$\eta_{\theta}(t)$, such that:
\begin{align}
    \langle \eta_{\theta}(t) \eta_{\theta'}(t') \rangle = 2 D \delta(t-t') \delta_{\theta\theta'},
\end{align}
where $\delta_{\theta\theta'}$ is the Kronecker delta.

Unfortunately, Eq.~\eqref{eq:discret_master_general} cannot be solved in closed form for arbitrary values of $\psi$, but simplifies in the two limits $\psi \to \infty$ (i.e. $z=\theta$, see section~\ref{Building a correlated discrete random field psi above one}) and $\psi \to 0$ (see section~\ref{Building a correlated discrete random field psi below one}). It will turn out that the latter limit allows us to calibrate the model with a single parameter, with the best goodness-of-fit over all other formulations.

\subsection{Building a correlated discrete random field when $\psi \gg 1$} \label{Building a correlated discrete random field psi above one}

In the limit $\psi \gg 1$, the change of variable accounting for psychological time vanishes (i.e. $z=\theta$). Eq.~\ref{eq:discret_master_general} becomes
\begin{equation}
    \label{eq:discret_master}
    \left\lbrace
    \begin{aligned}
    &\frac{{\rm d} A_\theta}{{\rm d} t}(t) = \frac{1}{\tau} \left[ -\mathcal{L}_d[A]_\theta(t) + \eta_\theta(t) \right], \\
    &A_{1}(t) - A_{0}(t)= 0,
    \end{aligned}
    \right.
\end{equation}
where the linear operator~$\mathcal{L}$ have been substituted by its naive discrete counterpart $\mathcal{L}_d$:
\begin{align}
    \mathcal{L}_d[A]_{\theta}(t) &:= A_{\theta}(t) - \frac{1}{\mu^2} \sum_{i=0}^{2}(-1)^{i}{\binom {2}{i}}A_{\theta+(1-i)}(t) \\ 
    &\quad + \frac{1}{\nu^4}\sum _{i=0}^{4}(-1)^{i}{\binom {4}{i}}A_{\theta+(2-i)}(t). \nonumber
\end{align}
This discrete operator~$\mathcal{L}_d$ mimics the impact of economic agents who compare the change of rate of a given tenor to the interpolation of the rates of its closest tenors. In fact, it seems intuitively plausible that agents primarily look at the two nearest tenors $\theta \pm 1$, corresponding to $1/\nu \to 0$. We will see below that the calibration of the model suggests that this is indeed the case.

\label{sec:discrete_d}
The solution to Eq.~\eqref{eq:discret_master} is given by
\begin{align}
    \label{eq:propagtor_model}
    A_\theta(t) =  \frac{1}{\tau} \int_{-\infty}^t {\rm d}t' \sum_{\theta'=0}^{+\infty} G_{\theta\theta'}(t-t') \eta_{\theta'}(t'),
\end{align}
where $G_{\theta\theta'}(t-t')$ is the \textit{propagator} of the noise $\eta_{\theta'}$ defined by 
\begin{align}
    & G_{\theta\theta'}(t-t') := \nonumber \\
    & \frac{1}{2\pi}\int_{-\pi}^{\pi} {\rm d}\xi \left( e^{i\xi(\theta-\theta')} + e^{i\xi(\theta+\theta')} \right) e^{-\frac{L_{d}(\xi)}{\tau}(t-t')},
\end{align}
with~$L_{d}(\xi) = 1 + 2\frac{(1-\cos{\xi})}{\mu^2} + 4\frac{(1-\cos{\xi})^2}{\nu^4}$ denoting the Fourier transform of~$\mathcal{L}_d$. The derivation of this result is detailed in Appendix~\ref{Solution to the discretized master equation}.

A crucial characteristic of the noise field~$A(t,\theta)$ is its auto-covariance across time and space. For~$\tau$ approaching $0$, the auto-covariance of~$A$ is found to be given by
 \begin{equation}
    \langle A_{\theta}(t) A_{\theta'}(t') \rangle =
    \left\lbrace
    \begin{aligned}
    &0, \text{ if } |t - t'| \gg \tau, \\
    &\frac{D}{\tau}  \mathcal{D}_1(\theta,\theta'), \text{ if } t=t',
    \end{aligned}
    \right.
\end{equation}
where the quantity $\mathcal{D}_k(\theta,\theta')$ is defined by
\begin{align}
    \label{eq:D_k}
    \mathcal{D}_k(\theta,\theta') = \frac{1}{\pi}\int_{0}^{\pi}{\rm d}\xi \frac{2\cos{\xi \theta}\cos{\xi \theta'}}{[L_{d}(\xi)]^{k}}.
\end{align}
The coarse-grained cumulative sum of~$A$ over a time interval~$\Delta t \gg \tau$, defined as
\[ 
\Delta A(t) := \int_{t-\Delta t/2}^{t+\Delta t/2} A_{\theta}(u) \mathrm{d}u,
\]
exhibits a behavior similar to its infinitesimal counterpart. For~$\tau \to 0$, the auto-covariance of~$\Delta A$ is given by
 \begin{equation}
    \label{eq:noise_correlator_linear}
    \langle \Delta A_{\theta}(t) \Delta A_{\theta'}(t') \rangle =
    \left\lbrace
    \begin{aligned}
    &0, \text{ if } |t-t'|> \Delta t, \\
    &2D \Delta t \, \mathcal{D}_2(\theta,\theta'), \text{ if } t=t'.
    \end{aligned}
    \right.
\end{equation}
Therefore, for~$\tau \ll 1$, both the infinitesimal and cumulative sum of~$A$ demonstrate martingale properties along the time axis while manifesting structured correlations across the spatial dimension~$\theta$. The proofs of these properties are provided in Appendix~\ref{Noise correlators psi above one}.

Following \cite{BaaquieBouchaud-2004}, it is tempting to still account for psychological time by replacing $\theta$ by $z(\theta)$ in Eq.\eqref{eq:noise_correlator_linear}, i.e.
\begin{align}
    \label{eq:noise_correlator_BBD3}
    \langle \Delta A_{\theta}(t) \Delta A_{\theta'}(t) \rangle = 2D \Delta t\mathcal{D}_2(z(\theta),z(\theta')).
\end{align}
Section~\ref{Modeling forward rates} shows that the equal-time Pearson correlation between the forward rates variations~$\Delta f_{\theta}(t)$ and~$\Delta f_{\theta'}(t)$ can be expressed thanks to the equal-time auto-covariance of~$\Delta A$. We will refer to the forward rate correlation model using Eq.~\ref{eq:noise_correlator_BBD3} as BBD3, for Baaquie-Bouchaud Discrete, three parameters.

Interestingly, one can derive closed-form formulas for~$\mathcal{D}_2(\theta,\theta')$ and~$\mathcal{D}_1(\theta,\theta')$ when~$(\theta,\theta') \in \mathbb{N}^2$, (see appendix~\ref{Closed formulas for the correlators}). However, we find that the numerical evaluation of integral~\eqref{eq:D_k} yields more stable results when extending the correlator $(k,k') \mapsto \mathcal{D}_2(k,k')$ to $\mathbb{R}^2$ as required by equation~\eqref{eq:noise_correlator_BBD3}.

\subsection{Building a correlated discrete random field when $\psi \ll 1$} \label{Building a correlated discrete random field psi below one}
If $\psi \ll 1$, the model can be written as a function of the products $\mu\psi$ and $\nu\psi$ only (see Appendix~\ref{Solution to the discretized Master Equation psi below one}). Hence, one can choose the parameter $\psi$ to be arbitrarily small to ensure that $\psi \ll 1 \leq \theta$. If we further consider that $\nu\psi  \gg 1$ -- which will turn out to be a reasonable assumption, as explained in section~\ref{A two-parameter version} -- the matrix $\mathcal{M}$ becomes:
\begin{align}
    \label{eq:curvy_M}
    \mathcal{M}_{\theta\theta'}= I - \frac{\theta}{\kappa^2}\left( I_{1} - I_{-1} \right)_{\theta\theta'} - \frac{\theta^2}{\kappa^2}\left(I_1 -2 I + I_{-1} \right)_{\theta\theta'}
\end{align}
where $\kappa := \mu\psi$.  Let $\mathcal{J}$ denote a matrix defined by
\begin{equation}
    \mathcal{J}_{\theta\theta'} =
    \left\lbrace
    \begin{aligned}
    &2, \text{ if } \theta=\theta'=0, \\
    &1, \text{ if } \theta=\theta'>0, \\
    &0, \text{ if } \theta\neq\theta'.
    \end{aligned}
    \right.
\end{equation}
In the limit $\psi \ll 1$ and $\nu \gg 1$, the solution to Eq.~\eqref{eq:discret_master_general} is given by
\begin{align}
    A(t) &= \frac{1}{\tau}\int_{-\infty}^{t} {\rm d}t' e^{-\frac{t-t'}{\tau}\mathcal{M}} \mathcal{J} \eta(t').
\end{align}
The derivation of this result is detailed in Appendix~\ref{Solution to the discretized Master Equation psi below one} in the general case of a finite $\nu$. Note that in the continuous time limit one recovers the BB model with a logarithmic psychological time.

We are interested in the auto-covariance across time and space of the noise field~$A(t,\theta)$. For~$\tau$ near $0$, Appendix~\ref{Noise correlators psi below one} shows that the autocovariance of $\Delta A$, the coarse-grain cumulative sum of $A(t)$ over the time interval $\Delta t \gg \tau$, is given by
\begin{equation}
    \label{eq:noise_correlator_BBDL}
    \langle \Delta A(t) \Delta A(t')^\top\rangle =
    \left\lbrace
    \begin{aligned}
    &0, \text{ if } |t-t'|> \Delta t, \\
    &2 D \Delta t \mathcal{M}^{-1}\mathcal{J}^2(\mathcal{M}^{-1})^\top, \text{ if } t=t'.
    \end{aligned}
    \right.
\end{equation}
We will refer to the forward rate correlation model using Eq.~\eqref{eq:noise_correlator_BBDL} as BBDL, for Baaquie-Bouchaud Discrete Logarithmic time, with only one parameter. Note that we have to invert numerically the matrix~$\mathcal{M}$ in Eq~\ref{eq:noise_correlator_BBDL} to compute forward rate correlations. To limit artificial deformations due to boundary effects for large $\theta$, we choose the dimension of the matrix $\mathcal{M}$ to be large ($\approx 500$) compared to the number of quoted forward rate contracts ($\approx 40$). This however has a minor impact on the final results.

\FloatBarrier
\section{Modeling forward rates} \label{Modeling forward rates}
\subsection{Forward rate diffusion}
The noise field previously defined is now employed to describe the dynamics of the forward rates. The diffusion equation for the variations of the forward rate, denoted as~${\rm d}f_{\theta}(t)$, is expressed as
\begin{align}
    \frac{{\rm d} f_{\theta}}{{\rm d} t} (t) = \gamma_{\theta}(t) + \sigma_{\theta} \widetilde{A}_{\theta}(t),
\end{align}
where the drift term $\gamma_\theta$ is set to zero, as discussed above, and 
\[ 
\widetilde{A}_{\theta}(t) = \frac{A_{\theta}(t)}{\sigma_A}
\] 
is the correlated noise field normalized by $\sigma_A$ where $\sigma_A^2 := \frac{1}{\Delta t} \langle \Delta A_{\theta}(t) \Delta A_{\theta}(t) \rangle$. This normalization ensures that $\sigma_{\theta}^2$ is the variance of the noise term driving the forward rate $f_\theta$. Note that because of this normalization, the value of $D$ is immaterial and can be set arbitrarily. We however keep it explicitly in the following for clarity.  

Consequently, at the mesoscopic scale~$\Delta t \gg \tau$, the variance of the forward rate increments 
\[ 
\Delta f_{\theta}(t) := \int_{t-\Delta t/2}^{t+\Delta t/2} {\rm d}f_{\theta}(t')
\] 
is given by 
\begin{align}
\langle \Delta f_{\theta}(t)^2 \rangle = \sigma_{\theta}^2 \Delta t.
\end{align}
Here we have assumed that the volatility of the infinitesimal forward rate variation is constant across time. The same formulas can be derived when considering constant per piece volatility on each day of length~$\Delta t$.

Finally, the equal-time Pearson correlation between the forward rates variations~$\Delta f_{\theta}(t)$ and~$\Delta f_{\theta'}(t)$ reads
\begin{align}
    \label{eq:forward_rate_correlation}
    \rho_{\theta\theta'} = \frac{\langle \Delta A_{\theta}(t) \Delta A_{\theta'}(t) \rangle}{\sqrt{\langle \Delta A_{\theta}(t) \Delta A_{\theta}(t) \rangle \langle \Delta A_{\theta'}(t) \Delta A_{\theta'}(t) \rangle}}.
\end{align}
It is clear from Eq.~\eqref{eq:forward_rate_correlation} that forward rate correlations can be expressed as a function of the noise field correlations $\langle \Delta A_{\theta}(t) \Delta A_{\theta'}(t) \rangle$. The calibration of our models in the following section is performed using Eq.~\eqref{eq:forward_rate_correlation} for different definitions of the noise field correlator (Eq.~\eqref{eq:noise_correlator_BBD3} or Eq.~\eqref{eq:noise_correlator_BBDL}).

\section{Calibration on correlation surfaces}
\subsection{Data}
\begin{figure}
    \centering \includegraphics{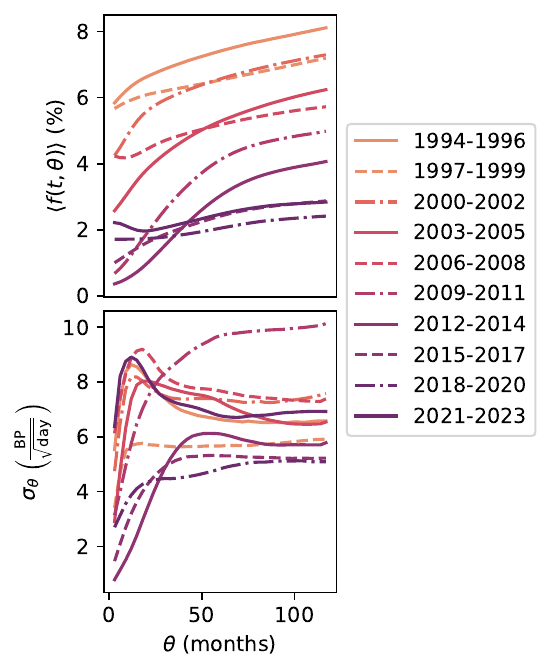}
    \caption{Unconditional mean and volatility of forward rates per tenor~$\theta$ across each of the three-year periods in our sample.}
    \label{fig:average_and_vol_midsize_periods}
\end{figure}

We interpret the instantaneous forward rate~$f(t,\theta)$ as the mid-price at time~$t$ of a $3$-month SOFR future contract maturing at~$t+\theta$. Our SOFR dataset comprises historical daily variations of these contracts' prices from $1994$ to $2023$, covering tenors from $3$ months to $117$ months. Consequently, we observe up to $n=39$ different tenors, resulting in $702$ distinct points in the correlation matrix (excluding the trivial diagonal points). $3$-month SOFR futures contracts were not available before March 2022, thus, prior to this, we used Eurodollar contracts. We present in Fig.~\ref{fig:average_and_vol_midsize_periods} the unconditional mean and volatility of forward rates per tenor~$\theta$ across each of the three-year periods in our sample. \citet{BouchaudEtAl-1999,BouchaudPotters-2003} observed that the mean of the FRC is a concave function of the tenor which is well approximated by a square-root, i.e. $\langle f(t,\theta) - f(t,0) \rangle \propto \sigma_1 \sqrt{\theta}$. This pattern suggests the forward rate~$f(t,\theta)$ can be interpreted as the probable adverse move that lenders could be facing at time $t+\theta$, since the pre-factor $\sigma_1$ matches quite well with the volatility of the short term rate $f(t,0)$ -- see \cite{MataczBouchaud-2000, MataczBouchaud-2000a} for a more detailed discussion. We confirm such behaviour for some of the three-year periods in our sample (see the top chart of Fig~\ref{fig:average_and_vol_midsize_periods}). 

Moreover, \citet{AminMorton-1994, HullWhite-1994, BouchaudEtAl-1999,BouchaudPotters-2003, FabozziMann-2005} have documented the humped shape of the volatility of the FRC. We do observe a peak in volatility~$\sigma_{\theta}$ around $\theta=12$ months for several of the three-year periods in our sample (see the bottom chart of Fig~\ref{fig:average_and_vol_midsize_periods}). This was interpreted in the same adverse move spirit as above, as the recent trend of the short term rate extrapolated in the future, see again see \cite{MataczBouchaud-2000, MataczBouchaud-2000a} for more on this point.

\begin{figure}
\centering
 \includegraphics[width=\linewidth]{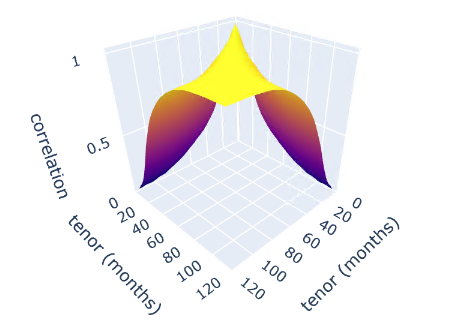}
 \caption{Empirical Pearson correlation surface between the daily forward rate increments of different tenors $\theta, \theta'$ for the period $2021-2023$. Notice that the surface is smooth across the diagonal $\theta =\theta'$, with a curvature that decreases with $\theta.$} 
 \label{fig:3dplot_correlation_prices_screen_shot}
\end{figure}

Central to the present study, the empirical Pearson correlations~$\hat{\rho}_{\theta\theta'}$ among the daily forward rate increments of tenor $\theta$ and $\theta'$ are depicted in Fig.~\ref{fig:3dplot_correlation_prices_screen_shot} for the specific period~$2021-2023$. As noted by \citet{BaaquieBouchaud-2004}, these correlations form a very smooth surface -- necessitating the introduction, within a continuous $\theta$ model, of a stiffness term in Eq.~\eqref{eq:action}, without which this surface would exhibit a cusp singularity along the diagonal. Note furthermore that the curvature along the diagonal decreases as the tenor $\theta$ increases which, as already alluded to, did motivate the introduction of a perceived, ``psychological time''. 

\subsection{Calibration over the whole sample} \label{Calibration over the whole sample}
We fit our micro-founded three parameter discrete model BBD3 (using Eq.~\eqref{eq:noise_correlator_BBD3}) to the observed correlation matrix over the period $1994-2023$, defining our whole sample. For this purpose, we define the error variance $\Sigma^2$ by
\begin{align}
    \Sigma^2 := \frac{1}{n^2}\sum_{\theta,\theta'} \left(\mathcal{E}_{\theta\theta'} - \frac{1}{n^2}\sum_{\theta,\theta'} \mathcal{E}_{\theta\theta'}\right)^2,
\end{align}
where $\mathcal{E}_{\theta\theta'} = \rho_{\theta\theta'} - \hat{\rho}_{\theta\theta'}$ is the difference between theoretical and empirical correlations for the forward rates of tenor $\theta$ and $\theta'$. We will refer to $\Sigma$ as the typical error of fit. We will use this indicator to assess the accuracy of our model.

The minimization of the error variance $\Sigma^2$ yields an optimal set of parameters~${\bf{p}}^* = (\psi^*,\mu^*, \nu^*)$ for the period $1994-2023$. The results in Fig.~\ref{fig:largest_anti_diag_['BBL3', 'BBD3', 'BBD2', 'BBDL']_full_period} represent such a fit along the largest anti-diagonal direction and gives the typical error over the whole surface, showing the high accuracy of the BBD3 model. We find the optimal parameters $\psi^*=2.06$ months, $\mu^*=1.06$, and $\nu^*=2.21$, for which $\Sigma = 1.47 \%$.

\begin{figure}
    \centering
    \includegraphics{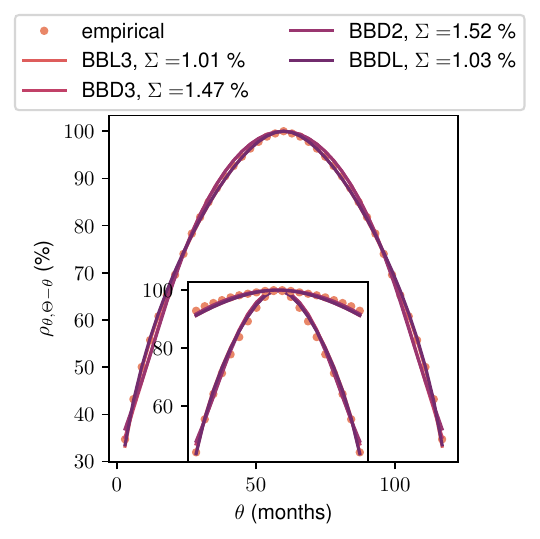}
    \caption{Dots represent the empirical correlation $\rho_{\theta\theta'}$ along the longest stretch perpendicular to the diagonal, i.e. $\theta' = \Theta - \theta$ , where $\Theta$ is the maximum available maturity. The plain lines are the best fit for the period $1994-2023$ in the continuous regularized model BBL3 (Eq.~\eqref{eq:BBL}) and our micro-founded discrete models: BBD3 (Eq.~\eqref{eq:noise_correlator_BBD3}), BBD2 (Eq.~\eqref{eq:noise_correlator_BBD3} with $\nu \to \infty$), and BBDL (Eq.~\eqref{eq:noise_correlator_BBDL}). The inset represent empirical and fitted correlations along two other anti-diagonals, defined by $\theta' = \frac{1}{2}\Theta - \theta$ and $\theta' = \frac{3}{2}\Theta - \theta$ respectively. }
    \label{fig:largest_anti_diag_['BBL3', 'BBD3', 'BBD2', 'BBDL']_full_period}
\end{figure}

\subsection{A two-parameter version} \label{A two-parameter version}

The interpretation of the discrete model in terms of a mean reverting force driving back tenor $\theta$ to the average of its two nearby tenors $\theta \pm 1$ suggests that the discrete fourth order derivative may in fact not be needed, since the discrete second order derivative effectively leads to a fourth order term in the continuous limit. In other words, one can set $\nu = \infty$ without creating a ``kink'' in $\rho_{\theta \theta'}$. This reduces the number of parameters to just two, $\psi$ and $\mu$, a version of the model that we will call BBD2. 

The calibration of BBD2 fully vindicates the above intuition: we find that the optimal values of the parameters over the full sample are given by $\psi^*=2.00$ months and $\mu^*=1.01$, corresponding to a typical error $\Sigma$ of $1.52$ \%, only $5$ basis points larger than to the one found for BBD3 with one less parameter (see Fig.~\ref{fig:largest_anti_diag_['BBL3', 'BBD3', 'BBD2', 'BBDL']_full_period}).

\begin{figure}
    \centering
    \subcaptionbox{Eigenvalues $\lambda_{1,2}$}{\includegraphics[width=0.38\linewidth]{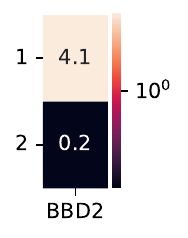}}
    \subcaptionbox{Eigenvectors ${\bf{e}}_{1,2}$}{\includegraphics[width=0.55\linewidth]{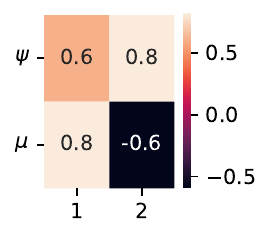}}
    \caption{Eigenvalues $\lambda$ and eigenvectors ${\bf{e}}$ of the Hessian matrix at ${\bf{p}}^*$ in the BBD2 model.}
    \label{fig:Hessian_eigen_values_['BBD2']}
\end{figure}

To analyze the structure of our two-parameter model, we study the sensitivity of the calibration error to each parameter. For this purpose, we define the Hessian matrix $\mathcal{H}$ as the second-order derivative of the typical error $\Sigma$, computed at the optimal set of parameters ${\bf{p}}^*$, i.e.
\begin{align}
   \mathcal{H}_{ij} := p^*_i p^*_j \left. \frac{\partial^2 \Sigma}{\partial p_i \partial p_j} \right|_{p_i=p^*_i, p_j=p^*_j}
\end{align}
The eigenvalues $\lambda = (\lambda_1,\lambda_2)$ and eigenvectors $(\bf{e}_1,\bf{e}_2)$ of the Hessian matrix $\mathcal{H}$ are presented in Fig.~\ref{fig:Hessian_eigen_values_['BBD2']} for the BBD2 model. It appears that $\lambda_1 \gg \lambda_{2}$, which means that only the combination of parameters along the ${\bf{e}}_1$ direction is relevant, the other direction being ``sloppy'' (see for example \citet{BrownSethna-2003,WaterfallEtAl-2006,GutenkunstEtAl-2007}). Fig.~\ref{fig:Hessian_eigen_values_['BBD2']} reveals that the main sensitivity mode of the BBD2 model is the a-dimensional product of parameters $\kappa := \psi \times \mu$. This suggests that we can further reduce the number of parameters to just one, as shown by the calculations of section~\ref{Building a correlated discrete random field psi below one} and successfully explored in the next section.

\subsection{A one-parameter version} \label{A one-parameter version}
The formulation of the model as a non-linear differential equation shows the line tension $\mu$ and psychological time parameter $\psi$ play similar roles. The line tension sets the weight on the second derivative while the psychological time defines the distance $ {\rm d}\theta$ between two consecutive tenors (see Eq.~\ref{eq:master_general}). Section~\ref{Building a correlated discrete random field psi below one} shows that these two parameters actually collapse as soon as $\psi$ is sufficiently small. This reduces the number of parameters to just $\kappa$, a version of the model previously mentioned as BBDL.

The calibration of BBDL outperforms all other tested models. We find that the optimal value for $\kappa$ is $0.92$, corresponding to a typical error $\Sigma$ of $1.03\%$. This time, the error is $44$ basis points lower than to the one found for BBD3 which has two more parameters, but is unable to explore the regime probed by BBDL (see Fig.~\ref{fig:largest_anti_diag_['BBL3', 'BBD3', 'BBD2', 'BBDL']_full_period}). In view of this, we focus on the more parsimonious version BBDL in the following, where we calibrate the model independently on each three-year sub-period.

\subsection{BBDL calibration on separate three-year periods} \label{BBDL calibration on separate three-year periods}
\begin{figure}
    \centering
    \includegraphics[width=0.9\linewidth]{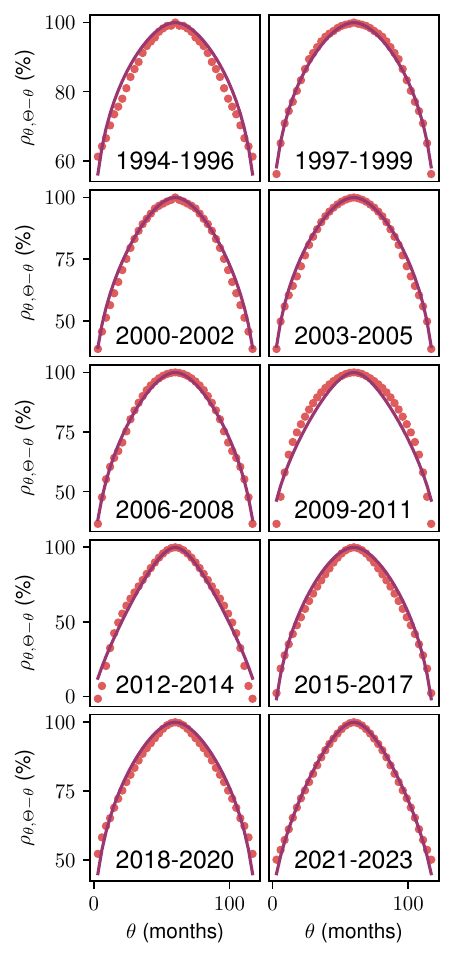}
    \caption{Dots represent the empirical correlation $\rho_{\theta\theta'}$ along the longest stretch perpendicular to the diagonal, i.e. $\theta' = \Theta - \theta$ , where $\Theta$ is the maximum available maturity. The plain lines are the best fit for our micro-founded discrete model BBDL (using Eq.~\eqref{eq:noise_correlator_BBDL}). It is clear from these plots that BBDL provides very accurate fits for all sub-periods, except 2009-2014.}
    \label{fig:largest_anti_diag_['BBD2']_all_periods}
\end{figure}

\begin{figure}
    \centering
    \includegraphics[width=0.9\linewidth]{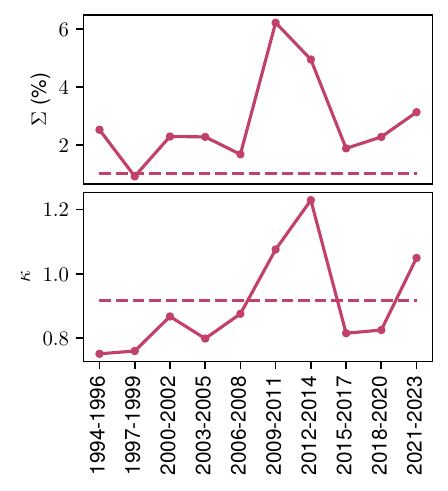}
    \caption{Optimal typical error and fitted parameter $\kappa$ obtained with our one-parameter micro-founded discrete model BBDL, Eq.~\eqref{eq:noise_correlator_BBDL}. The dotted lines corresponds to the calibration results on the $1994-2023$ period. Note that the typical error $\Sigma$ yielded by the fit on the whole sample is not equal to the average error across sub-periods.}
    \label{fig:parameters_all_periods_['BBD2']}
\end{figure}

Fig.~\ref{fig:largest_anti_diag_['BBD2']_all_periods} represents the largest anti-diagonal of the fit of the BBDL model over the whole correlation surface, for each three-year sub-period. The typical error and calibrated parameters for these intervals, detailed in Figure~\ref{fig:parameters_all_periods_['BBD2']}, demonstrate the stability of the parameter $\kappa$  throughout the assessed periods. However, periods characterized by significant monetary policy shifts exhibit a decrease in line tension (reflected in a higher value of $\kappa$) and goodness-of-fit (higher typical error~$\Sigma$). Specifically, three periods present line tension lower than its long-term value: (i) the $2009-2011$ span, during the first and second rounds of Quantitative Easing, with the Federal Reserve purchasing $900$ billion dollars in US treasury bonds; (ii) the $2012-2014$  phase, with an additional $800$ billion in bond purchases (third Quantitative Easing); and (iii) the $2021-2023$ period, notable for the fourth Quantitative Easing amid the COVID-19 pandemic. This suggests that asset purchase programs induce periods of heightened curvature on the forward rate curve, i.e. more decoupling between nearby tenors.


\subsection{Comparison with \citet{BaaquieBouchaud-2004}}
In this section we perform a comparison of the accuracy and stability of all the tested models with the approach initially proposed by \citet{BaaquieBouchaud-2004}. For the ease of convenience, Table~\ref{tab:Models' definition} in appendix~\ref{Models' definition} summarizes the definition of these models.

We show in Fig.~\ref{fig:largest_anti_diag_['BBL3', 'BBD3', 'BBD2', 'BBDL']_full_period} that the continuous Baaquie-Bouchaud model with logarithmic psychological time (BBL3), see Eq.~\eqref{eq:BBL} achieves the same global accuracy with $\Sigma = 1.01 \%$, compared to $1.03 \%$ for BBDL. However, removing only one of its three parameters now considerably degrades the goodness-of-fit. For example, removing the stiffness term increases the typical error $\Sigma$ from $1.01\%$ to $4.06\%$. As obvious from Appendix~\ref{Two-parameters versions}, Fig. \ref{fig:argest_anti_diag_['BBL2', 'BBD2']_full_period}, this is chiefly because the correlation surface develops a cusp around the diagonal $\theta=\theta'$, which was actually the very reason why \citet{BaaquieBouchaud-2004} introduced such a stiffness term!

The discrete BBDL model therefore appears superior not only because it is micro-founded and intuitively compelling, but also because it is more parsimonious: it naturally gets rid of the diagonal cusp without having to introduce any additional parameter. The discrete second derivative $f_{\theta+1} + f_{\theta-1} - 2f_{\theta}$ indeed formally contains continuous derivatives $\partial_\theta^{2k} f(\theta)$ of all even orders, and is therefore sufficient to regularize the correlation function across the diagonal. Furthermore, the psychological time parameter $\psi$ merges, in the limit $\psi \to 0$, with the line tension parameter $\mu$, if the change of variable is encoded in the dynamical master equation~\eqref{eq:master_general}. Indeed, intuitively, changing the distance between tenors ($\psi$) or the line tension ($\mu$) is equivalent.

Lastly, we compare the stability of all the tested models over successive three-year intervals within our dataset. Fig.~\ref{fig:parameters_all_periods_method_direct_['BBL3', 'BBD3', 'BBD2', 'BBDL']} displays the typical error and calibrated parameters across all periods, illustrating a consistent fit quality except during significant monetary policy shifts (refer to Section~\ref{BBDL calibration on separate three-year periods}). However the continuous and discrete three-parameter models (BBL3 and BBD3) demonstrate a high parameter instability across periods (note the log scale for $\mu$ and $\nu$). The more parsimonious approaches (BBD2 and BBDL), with only a slightly higher level of error on some periods, exhibit significantly more stable parameters. Notably, BBDL emerges as the most stable model, with its single parameter $\kappa$ ranging from $0.8$ to $1.2$ across all periods.

\begin{figure}
    \centering
    \includegraphics[width=\linewidth]{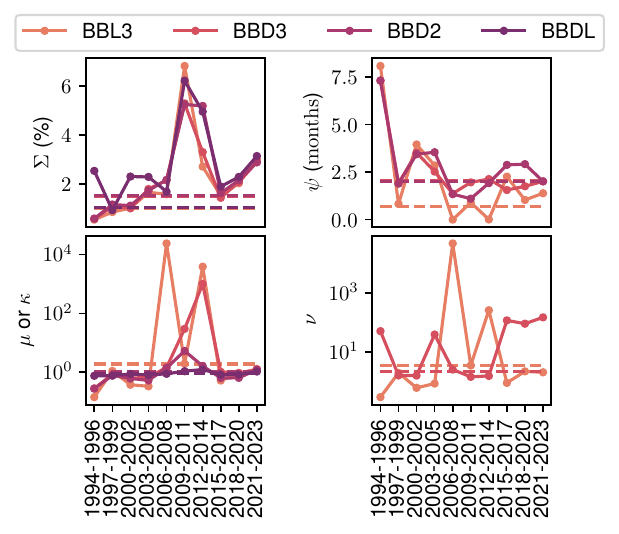}
    \caption{Typical error and fitted parameters within the continuous regularized model BBL3 (Eq.~\eqref{eq:BBL}) and our micro-founded discrete models: BBD3 (Eq.~\eqref{eq:noise_correlator_BBD3}), BBD2 (Eq.~\eqref{eq:noise_correlator_BBD3} with $\nu \to \infty$), and BBDL (Eq.~\eqref{eq:noise_correlator_BBDL}). The dotted lines correspond to the calibration results for each model over the $1994-2023$ period. The bottom left figure shows the parameter $\kappa$ for the BBDL model or the parameter $\mu$ for all the others. While $\mu$ (or $\kappa$) and $\nu$ are dimensionless in the discrete models, these parameters are expressed in $\text{3 months}^{-1}$ in the case of the BBL3 model. Table~\ref{tab:Models' definition} in appendix~\ref{Models' definition} summaries the definition of the tested models.}
    \label{fig:parameters_all_periods_method_direct_['BBL3', 'BBD3', 'BBD2', 'BBDL']}
\end{figure}

\section{The Epps effect}

An important vindication of our framework, in particular the {\it dynamical} construction of the field $A_\theta$ using Eq. \eqref{eq:principle}, is our ability to account for the so-called Epps effect \cite{Epps-1979} in a very natural way. Indeed, the auto-covariance of~$\Delta A/\sqrt{\Delta t}$ is predicted by the theory to increase from $0$ for~$\Delta t \to 0$ at fixed~$\tau$, to~$2D  \, \mathcal{D}_2(\theta,\theta')$ when $\Delta t \gg \tau$ (see Eq.~\eqref{eq:correlator_all_time_scales} in appendix~\ref{Noise correlators psi above one} for the case $\psi \gg 1$ and Eq.~\eqref{eq:correlator_all_time_scales_psi_below_one} in appendix~\ref{Noise correlators psi below one} for the case $\psi \ll 1$). 

\begin{figure}
    \centering
    \includegraphics{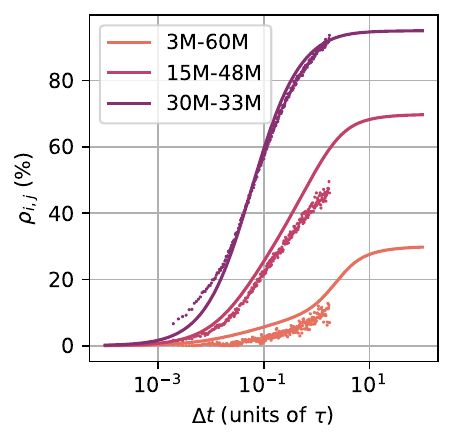}
    \caption{Plain lines: theoretical Pearson correlation coefficients among three pairs of forward rate variations~$(\Delta f_\theta, \Delta f_{\theta'})$ as a function of the time scale~$\Delta t$ (see~\eqref{eq:correlator_all_time_scales_psi_below_one}). Using the empirical correlations of the pair $30$-$33$ months, the parameter~$\varepsilon$ of the idiosyncratic white noise was calibrated to $1.6 \times 10^{-3}$, and the characteristic time of the Epps effect $\tau$ to $36$ minutes. This figure also shows the theoretical correlations yielded by this set of parameters for two other pairs ($15$-$48$ and $3$-$60$ months).
    Dots: empirical Pearson correlation coefficients for three pairs of SOFR Futures prices for the year $2021$ at time scales ranging from $4$ seconds to one hour.}
    \label{fig:epps_theo_bb_num_mu}
\end{figure}

Without modifying our model at the daily time scale, we may postulate that an additional, small white noise contributes to $A_{\theta}$, originating for example from the idiosyncratic dynamics of the order flow, or from the bid-ask bounce. We shall assume that the variance of such a noise is $2D \varepsilon\Delta t$, with $\varepsilon$ an extra $\theta$-independent parameter such that 
\begin{align}
    \varepsilon \ll \mathcal{C}(\theta,\theta),
\end{align}
where $\mathcal{C}(\theta,\theta')$ is the auto-covariance the noise field $\Delta A/\sqrt{2D\Delta t}$ without idiosyncratic noise for $\Delta t \gg \tau$ \footnote{More precisely, in the case $\psi \gg 1$, $\mathcal{C}(\theta,\theta') = \mathcal{D}(z(\theta),z(\theta'))$, and when $\psi \ll 1$, $\mathcal{C}(\theta,\theta') = (\mathcal{M}^{-1}\mathcal{J}^2(\mathcal{M}^{-1})^\top)_{\theta\theta'}$.}. We then obtain the following scale-dependent covariance structure for~$\Delta A$:
\begin{equation}
    \label{eq:theo_epps}
    \langle \Delta A_{\theta}(t) \Delta A_{\theta'}(t) \rangle =
    \left\lbrace
    \begin{aligned}
    &2D \varepsilon \, \delta_{\theta\theta'} \, \Delta t + O(\Delta t^2) &\text{ for } \Delta t \ll \tau, \\
    & 2D \mathcal{C}(\theta,\theta') \, \Delta t &\text{ for } \Delta t \gg \tau,
    \end{aligned}
    \right.
\end{equation}
that we can compare with empirical data.
Each plain line in Fig.~\ref{fig:epps_theo_bb_num_mu} represents the correlation $\rho_{\theta \theta'}$ across different time scales~$\Delta t$ among pairs of forward rate variations $(\Delta f_\theta, \Delta f_{\theta'})$, as given by our model in the case $\psi \gg 1$ (see Eq.~\eqref{eq:correlator_all_time_scales_psi_below_one}) calibrated on daily correlations (cf. section~\ref{Calibration over the whole sample}) with an additional fitting parameter~$\varepsilon$.\footnote{A natural extension would be to let $\varepsilon$ depend on $\theta$. A rank one specification, for example, would read $\varepsilon_{\theta \theta'} = (C_1 + C_2 \log \theta)(C_1 + C_2 \log \theta')$. We have not tried to calibrate such a model, since the simplest version $C_2=0$ is satisfactory for our purpose.}   Fig.~\ref{fig:epps_theo_bb_num_mu} clearly demonstrates that our model is able to reproduce the whole dependence of the empirical correlations of pairs of SOFR Futures binned at different time scales (dots in Fig.\ref{fig:epps_theo_bb_num_mu}, see also Fig.~\ref{fig:df_epps_corr}). In fact, one can back out from this exercice the correlation time scale $\tau$ through a minimization of the differences between empirical and theoretical correlations across time scales. For the pair $30$-$33$ months, this leads to a very reasonable value $\tau \approx 36$ minutes that can be interpreted as the information propagation time along the FRC. This calibration also yields an optimal value for the size of the idiosyncratic noise $\varepsilon \approx 1.6 \times 10^{-3}$, which is an order of magnitude smaller than $\mathcal{C}(\theta,\theta')$ (ranging from $0.02$ to $4$ when $\psi \ll 1$). In addition, Fig.~\ref{fig:epps_theo_bb_num_mu} shows the theoretical correlations yielded by $\tau = 36$ minutes and $C = 1.6 \times 10^{-3}$ for two other pairs ($15$-$48$ and $3$-$60$ months). It indicates that a similar set of calibrated parameters would be obtained if we had used a different pair.

One could have used Eq.~\ref{eq:correlator_all_time_scales} in the case $\psi \gg 1$ to generate the Epps effect. It yields a similar shape of the correlation $\rho_{\theta \theta'}$ across different time scales~$\Delta t$ among pairs of forward rate variations (see Fig~\ref{fig:epps_theo_imf_2_log} in Appendix~\ref{Epps effect psi below one}). Using this approach, we find a larger $\varepsilon \approx 0.026$, consistently with higher value of $\mathcal{C}(\theta,\theta')$  when $\psi \gg 1$ (in the range of $0.14$ to $0.48$) and a very similar value for $\tau \approx 21$ minutes. 

\section{Conclusion}
In this paper, we have reformulated the forward interest rate field theory of \citet{BaaquieBouchaud-2004} to account in a unified manner for two important features: (a) the discrete set of traded maturities and (b) the scale dependent structure of the correlation matrix across maturities (the Epps effect). Both points are related to market mechanisms underlying our modeling assumptions. 

Indeed, we believe that the emergent correlation structure is a result of market participants reacting to high frequency shocks affecting the different tenors along the forward rate curve, which get corrected in time and transmitted along maturities through a self-referential mechanism. Intuitively, the dynamics of rates maturing at $t+\theta$ in the future cannot be decoupled from rates maturing at $t + \theta'$ when $|\theta-\theta'|$ is small. This is encoded, within our framework, via relative mean-reverting forces proportional to the discrete Laplacian and discrete fourth derivative of the returns along the maturity axis. As it turns out, the discrete fourth derivative plays a minor role and can be neglected -- whereas this term was crucial in the continuous time version of \citet{BaaquieBouchaud-2004}.

We have shown that such a parsimonious specification, further equipped with the notion of ``psychological time'' that shrinks the perceived distance between far away maturities, allows one to reproduce remarkably well (with an error around $1\%$) the full correlation structure of the forward rate curve, in particular the maturity dependent curvature of the correlation perpendicular to the diagonal $\theta=\theta'$. The single parameter of the model is found to be particularly stable across all the tested periods. Quite remarkably, we find that the data is compatible with the assumption of a logarithmic dependence of the perceived time as a function of real time, which translates into the hyperbolic discounting factor advocated in the behavioral economics literature \citep{FarmerGeanakoplos-2009}. From our calibration to the data, we find that the cross-over time between normal time flow and logarithmic time flow occurs around 2 months in the future. This also means that a year ten years from now is perceived by the bond markets as one week in real time. This is quite an extreme distortion of future time that reflects the extremely myopic nature of financial markets.  

Finally, our approach also quantitatively reproduces the empirical finding of negligible correlations at high frequencies \citep{Epps-1979}, which slowly build up at lower frequencies, see Fig. \ref{fig:epps_theo_bb_num_mu}, with a characteristic time scales on the order of 30 minutes. 
The modeling framework we advocate also captures several phenomena consistent with the market micro-structure literature, including (i) non-martingality of prices at short time scales and (ii) price-impact and cross-impact effects. These phenomena that will be detailed in a companion paper \citet{LeCozEtAl-2024}.

\section*{Acknowledgments}
We would like to express our gratitude to Michael Benzaquen, Damien Challet, Iacopo Mastromatteo  and Samy Lakhal, who contributed to our research through fruitful discussions. We also thank Bertrand Hassani and the ANRT (CIFRE number 2021/0902) for providing us with the opportunity to conduct this research at Quant AI Lab. Finally, we thank the referees for useful comments. 

This research was conducted within the Econophysics \& Complex Systems Research Chair, under the aegis of the Fondation du Risque, the Fondation de l’École polytechnique, the École polytechnique and Capital Fund Management.

\section*{Disclosure of interest}
The authors declare that there are no conflicts of interest to disclose.

\FloatBarrier
\newpage
\bibliographystyle{apsrev4-2} 
\bibliography{zotero}

\newpage
\appendix

\section{Notations} \label{Notations}
Table~\ref{tab:notations} summaries the notations used in this study.
\begingroup
\squeezetable
\begin{table}[h]
    \begin{ruledtabular}
    \begin{tabular}{p{0.14\linewidth} p{0.84\linewidth}}
    \multicolumn{1}{c}{Expression} & \multicolumn{1}{c}{Definition} \\
    \hline
    $n$ & The number of available SOFR futures. \\
    $t$ & The current time. \\
    $T$ & The maturity.\\
    $P(t,T)$ & The price at time t of a zero-coupon bond maturing at~$T$. \\
    $\theta$ & The time-to-maturity or tenor, in units of 3 months. \\
    $f(t,\theta)$ & The value at time t of the instantaneous forward rate of tenor~$\theta$ (continuous notation). \\
    $f_\theta(t)$ & The value at time t of the instantaneous forward rate of tenor~$\theta$ (discrete notation). \\
    $A(t,\theta)$ & The driftless correlated noise field (continuous notation).\\
    $A_\theta(t)$ & The driftless correlated noise field (discrete notation). \\
    $\eta(t,\theta)$ & The two-dimensional white noise on continuous space. \\
    $\eta_{\theta}(t)$ & The discrete white noise of tenor~$\theta$. \\
    $\sigma(t,\theta)$ & The volatility at time~$t$ of the infinitesimal variation of the instantaneous forward rate of time-to-maturity~$\theta$ (continuous notation). \\
    $\sigma_\theta(t)$ & The volatility at time~$t$ of the infinitesimal variation of the instantaneous forward rate of time-to-maturity~$\theta$ (discrete notation). \\
    $\gamma(t,\theta)$ & The drift at time~$t$ of the infinitesimal variation of the instantaneous forward rate of time-to-maturity~$\theta$ (continuous notation). \\
    $z(\theta)$ & The psychological time. \\
    $\psi$ & The psychological time parameter in the change of variable $z(\theta) = \psi \log\left(1 + \frac{\theta}{\psi}\right)$. \\
    $\bar{\psi}$ & The psychological time parameter in the change of variable $\bar{z}(\theta) = \theta^{\bar{\psi}}$ \citep{BaaquieBouchaud-2004}. \\
    $\mu$ & The line tension parameter. \\
    $\nu$ & The stiffness (or bending rigidity) parameter. \\
    $\kappa$ & Unique a-dimensional parameter in the BBDL model defined by $\kappa := \mu \psi$. \\
    $\mathcal{D}_{BB}(\theta,\theta')$ & The Baaquie-Bouchaud correlator \citep{BaaquieBouchaud-2004}. \\
    $\tau$ & The time scale for the emergence of correlations. \\
    $\Delta t$ & The time scale at which forward variations are observed, corresponding to one day unless specified otherwise. \\
    $\Delta \eta_\theta$ & The coarse-grained cumulative sum over the time scale $\Delta t$ of the two-dimensional white noise $\eta_\theta$. \\
    $\Delta A_\theta$ & The coarse-grained cumulative sum over the time scale $\Delta t$ of the correlated noise field $A_\theta$. \\
    $\Delta f_\theta$ & The forward rate increments over the time scale $\Delta t$. \\
    $\langle . \rangle$ & The average operator over the functional weight $e^{S[A]}$. \\
    $\delta(.)$ & The Dirac delta function. \\
    $\delta_{\theta\theta'}$ & The Kronecker delta. \\
    $\mathcal{L}[.] $ & The continuous linear differential operator on space. \\
    $\mathcal{O}[.] $ & The continuous non-linear differential operator on space. \\
    $\mathcal{L}_d[.] $ & The discrete linear differential operator on space. \\
    $\mathcal{M} $ & The discrete non-linear differential operator on space, using matrix notations. \\
    $L_d[.] $ & The Fourier transform of the discrete linear differential operator on space. \\
    $G_{\theta\theta'}(.) $ & Green function or \textit{propagator} of the discretized Eq.~\eqref{eq:discret_master} \\
    $\mathcal{D}_2(\theta,\theta')$ & The spatial correlator in the discrete BBD model. \\
    $\mathcal{D}_1(\theta,\theta')$ & The spatial correlator of the noise field. \\
    $I_k$ & A matrix with ones only on the $k$-th diagonal above the main diagonal. \\
    $I$ & The identity matrix. \\
    $\mathcal{J}$ & A diagonal matrix whose first entry is $2$ while all the other entries are ones. \\
    $2D\Delta t$ & The variance of $\Delta \eta$. \\
    $2D\epsilon\Delta t$ & The variance of the cumulative sum of the idiosyncratic two-dimensional white noise. \\
    $\mathcal{F}[f]$ & The Fourier transform of the function of the maturity~$\theta \mapsto f_\theta$. \\
    $\Sigma$ & The typical error between the empirical and the theoretical correlations.
    \end{tabular}
    \end{ruledtabular}
    \caption{Notations\label{tab:notations}}
\end{table}
\endgroup
\FloatBarrier

\section{Models' definition} \label{Models' definition}
Table~\ref{tab:Models' definition} summaries the definition of the models used in this study.
\begingroup
\renewcommand{\arraystretch}{2}
\begin{table}[h]
    \begin{ruledtabular}
    \begin{tabular}{p{0.1\linewidth} p{0.85\linewidth}}
    \multicolumn{1}{c}{Model} & \multicolumn{1}{c}{Noise correlator $\langle \Delta A_{\theta}(t) \Delta A_{\theta'}(t) \rangle$} \\
    \hline
    BB04 & $\mathcal{D}_{BB}(\theta^{\bar{\psi}},{(\theta')}^{\bar{\psi}})$ (see Eq.~\eqref{eq:BB04})\\
    BBL3 & $\mathcal{D}_{BB}(z(\theta),z(\theta'))$ (see Eq.~\eqref{eq:BB04})\\
    BBL2 & $\mathcal{D}_{BB}(z(\theta),z(\theta'))$ with $\nu \to \infty$ (see Eq.~\eqref{eq:BB04})\\
    BBD3 & $2D \Delta t\mathcal{D}_2(z(\theta),z(\theta'))$. (see Eq.~\eqref{eq:D_k})\\
    BBD2 & $2D \Delta t\mathcal{D}_2(z(\theta),z(\theta'))$ with $\nu \to \infty$ (see Eq.~\eqref{eq:D_k})\\
    BBDL & $2 D \Delta t \mathcal{M}^{-1}\mathcal{J}^2(\mathcal{M}^{-1})^\top$ with $\nu\psi \to \infty$ and $\psi \to 0$ (see Eq.~\eqref{eq:curvy_M}). \\
    \end{tabular}
    \end{ruledtabular}
    \caption{Models' definition\label{tab:Models' definition}}
\end{table}
\endgroup
\FloatBarrier

\section{Solution to the discretized Master Equation when $\psi \gg 1$} \label{Solution to the discretized master equation}
We define the \textit{propagator}~$\mathcal{G}_{\theta\theta'}(t,t')$ as the solution to
\begin{align}
    \label{eq:propagator}
    \frac{\partial \mathcal{G}}{\partial t} + \frac{1}{\tau} \mathcal{L}_d[\mathcal{G}] = \delta(t-t')  \delta_{\theta\theta'}.
\end{align}
The linear operator in Eq.~\eqref{eq:propagator} is translation invariant as it has constant coefficients. It indicates that $\mathcal{G}_{\theta\theta'}(t,t')$ depends only on $\theta-\theta'$ and $t-t'$. The symmetry of the functions~$\theta \mapsto \mathcal{L}_d[\mathcal{G}]_{\theta}(t)$ and~$\theta \mapsto \delta_{\theta\theta'}$ further ensures that the dependence of the propagator with respect to space depends only of the absolute value of the differences $|\theta-\theta'|$. Let $H$ denote the Heaviside function. Applying discrete Fourier decomposition to the dimension~$\theta-\theta'$ yields a particular solution:
\begin{equation}
    \mathcal{F}[\mathcal{G}](\xi,t-t') =  H(t-t') e^{-\frac{L_{d}(\xi) (t-t')}{\tau}},
\end{equation}
where $L_{d}(\xi) = 1 + 2\frac{(1-\cos{\xi})}{\mu^2} + 4\frac{(1-\cos{\xi})^2}{\nu^4}$ and $ \mathcal{F}[\mathcal{G}](\xi,t-t')$ denote the spatial Fourier transform of $\mathcal{L}_d[A]_{\theta}(t)$  and $\mathcal{G}_{\theta\theta'}(t-t')$ respectively. These functions are continuous in $\xi$ and $t$.

Seeking a solution on $\mathbb{R}\times\mathbb{Z}$ for the discretized Eq.~\eqref{eq:discret_master} without boundary conditions, we extend the noise $\eta$ so that $\eta_{\theta}(t) = 0$ for $\theta\in \mathbb{Z}_-^*$. The solution is
\begin{align}
    \label{eq:solution}
     \mathcal{A}_\theta(t) &= \frac{1}{\tau} \int_{\mathbb{R}} {\rm d}t' \sum_{\theta'=-\infty}^{+\infty} \mathcal{G}_{|\theta-\theta'|}(t-t') \eta_{\theta'}(t') \nonumber \\
     &=  \frac{1}{\tau} \int_{-\infty}^t {\rm d}t' \sum_{\theta'=0}^{+\infty} \mathcal{G}_{|\theta-\theta'|}(t-t') \eta_{\theta'}(t').
\end{align}

Similarly, $\mathcal{A}_{-\theta}(t)$ solves the discretized Eq.~\eqref{eq:discret_master} without boundary conditions when substituting $\eta_{\theta}(t)$ with $\eta_{-\theta}(t)$. Thus, a specific solution on $\mathbb{R}\times\mathbb{N}$ that satisfies the Neumann boundary condition is
\begin{equation}
    A_\theta(t) := \mathcal{A}_\theta(t) + \mathcal{A}_{-\theta}(t).
\end{equation}

The propagator~$G_{\theta\theta'}(t-t')$ associated with this solution is defined as
\begin{align}
    \label{eq:propagator_solution}
    &G_{\theta\theta'}(t-t') := \mathcal{G}_{|\theta-\theta'|}(t-t') + \mathcal{G}_{\theta+\theta'}(t-t') \nonumber \\
    &= \frac{1}{2\pi}\int_{-\pi}^{\pi} {\rm d}\xi \left( e^{i\xi(\theta-\theta')} + e^{i\xi(\theta+\theta')} \right) \mathcal{F}[\mathcal{G}](\xi,t-t').
\end{align}
For consistency with the centered discretization scheme, the inverse Fourier transform is centered on $[-\pi,+\pi]$.

\section{Noise correlators when $\psi \gg 1$} \label{Noise correlators psi above one}

\subsection{Autocovariance of the correlated noise}
For $\psi \gg 1$, the autocovariance of $A$ is defined by
\begin{align}
    &\langle A_{\theta}(t) A_{\theta'}(t') \rangle :=\frac{1}{\tau^2} \int_{-\infty}^t {\rm d}u \int_{-\infty}^{t'} {\rm d}v\\ \nonumber & \sum_{(U,V)\in \mathbb{N}^2} G_{\theta U}(t-u) G_{\theta' V}(t'-v) \left \langle \eta_{U}(u) \eta_{V}(v) \right \rangle.
\end{align}
Recalling that 
\begin{equation}
     \left \langle \eta_{U}(u) \eta_{V}(v) \right \rangle = 2D \delta(u-v) \delta_{UV}, 
\end{equation}
we derive
\begin{align}
    &\langle A_{\theta}(t) A_{\theta'}(t') \rangle \nonumber \\
    &= \frac{2D}{\tau^2} \int_{-\infty}^{t \wedge t'} {\rm d}u \sum_{U\in \mathbb{N}}  G_{\theta U}(t-u) G_{\theta'U}(t'-u).
\end{align}
Substituting the propagator $G$ with its expression in Eq.~\eqref{eq:propagator_solution} yields 
\begin{align}
    &\langle A_{\theta}(t) A_{\theta'}(t') \rangle \nonumber \\
    &= \frac{2D}{(2\pi\tau)^2} \iint_{-\pi}^{\pi} {\rm d}\xi {\rm d}\xi' \int_{-\infty}^{t \wedge t'} {\rm d}u e^{-\frac{1}{\tau}\left(L_{d}(\xi)(t-u)+L_{q'}(t'-u)\right)} 
    \nonumber \\ 
    & \hspace{1cm} \sum_{U\in \mathbb{N}} \left(e^{i\xi(\theta-U)}+e^{i\xi(\theta+U)}\right)\left(e^{i\xi'(\theta'-U)}+e^{i\xi'(\theta'+U)}\right) \nonumber \\
    &= \frac{2D}{(2\pi\tau)^2} \iint_{-\pi}^{\pi} {\rm d}\xi {\rm d}\xi' \int_{-\infty}^{t \wedge t'} {\rm d}u e^{-\frac{1}{\tau}\left(L_{d}(\xi)(t-u)+L_{d}(\xi')(t'-u)\right)} 
    \nonumber \\ 
    &  \hspace{3cm} e^{i(\xi\theta+\xi'\theta')} \sum_{U\in \mathbb{Z}} \left(e^{iU(\xi+\xi')}+e^{iU(\xi-\xi')}\right) \nonumber \\
    &= \frac{2D}{2\pi\tau^2} \int_{-\pi}^{\pi} {\rm d}\xi \int_{-\infty}^{t \wedge t'} {\rm d}u e^{-\frac{L_{d}(\xi)}{\tau}\left(t+t'-2u\right)}  \left( e^{i\xi(\theta-\theta')} + e^{i\xi(\theta+\theta')} \right), \nonumber
\end{align}
having noted that $L_{d}(-\xi)=L_{d}(\xi)$. The computation of the integral with respect to time gives the expression of the autocovariance of $A$:
\begin{align}
    &\langle A_{\theta}(t) A_{\theta'}(t') \rangle \nonumber \\
    &= \frac{D}{2\pi\tau} \int_{-\pi}^{\pi}{\rm d}\xi \frac{e^{i\xi(\theta-\theta')} + e^{i\xi(\theta+\theta')}}{L_{d}(\xi)} e^{-\frac{L_{d}(\xi)}{\tau}|t-t'|}.
\end{align}

We now define the quantity $\mathcal{D}_1(\theta,\theta')$ by
\begin{align}
     \mathcal{D}_1(\theta,\theta') = \frac{1}{2\pi}\int_{-\pi}^{\pi}{\rm d}\xi \frac{e^{i\xi(\theta-\theta')} + e^{i\xi(\theta+\theta')}}{L_{d}(\xi)}.
 \end{align}
$L_{d}(\xi)$ is symmetric, so one can reformulate the above quantity as 
\begin{align}
     \mathcal{D}_1(\theta,\theta') = \frac{1}{\pi}\int_{0}^{\pi}{\rm d}\xi \frac{2\cos{\xi \theta}\cos{\xi \theta'}}{L_{d}(\xi)}.
 \end{align}

Therefore, for $\tau$ close to $0$, the covariance of $A$ simplifies to
 \begin{equation}
    \langle A_{\theta}(t) A_{\theta'}(t') \rangle =
    \left\lbrace
    \begin{aligned}
    &0, \text{ if } t \neq t', \\
    &\frac{D}{\tau}  \mathcal{D}_1(\theta,\theta'), \text{ if } t=t'.
    \end{aligned}
    \right.
\end{equation}

In our case, $L_{d}(\xi) = 1 + 2\frac{1-\cos{\xi}}{\mu^2} + 4\frac{(1-\cos{\xi})^2}{\nu^4}$. Hence, one can show that, for $\xi \ll 1$ and $\tau \ll 1$, the autocovariance~$\langle A_{\theta}(t) A_{\theta'}(t') \rangle$ converges to the correlator~$ \delta(t-t') \mathcal{D}_{BB}(\theta,\theta')$ introduced by~\citet{BaaquieBouchaud-2004}.

\subsection{Autocovariance of the cumulative correlated noise}
We define the coarse-grained cumulative sum of~$A_{\theta}$ over a time interval~$\Delta t \gg \tau$ by
\begin{align}
    \Delta A_{\theta} = \int_{t-\Delta t/2}^{t+\Delta t/2}  A_{\theta}(u) {\rm d}u.
\end{align}
We derive the autocovariance of $\Delta A$ when $\psi \gg 1$,
\begin{align}
    & \langle \Delta A_{\theta}(t) \Delta A_{\theta'}(t') \rangle = \int_{t-\Delta t/2}^{t+\Delta t/2} {\rm d}u \int_{t'-\Delta t/2}^{t'+\Delta t/2}  {\rm d}v \langle A_{\theta}(u) A_{\theta'}(v) \rangle, \nonumber \\
    & = \frac{D}{2\pi\tau}\int_{-\pi}^{\pi}{\rm d}\xi\int_{t-\Delta t}^t {\rm d}u \int_{t'-\Delta t}^{t'} {\rm d}v  \nonumber \\
    & \hspace{3cm} \frac{e^{i\xi(\theta-\theta')} + e^{i\xi(\theta+\theta')}}{L_{d}(\xi)} e^{-\frac{L_{d}(\xi)}{\tau}|u-v|}.
\end{align}

We now define the quantity $\mathcal{D}_2(\theta,\theta')$ by
\begin{align}
     \mathcal{D}_2(\theta,\theta') &= \frac{1}{2\pi}\int_{-\pi}^{\pi}{\rm d}\xi \frac{e^{i\xi(\theta-\theta')} + e^{i\xi(\theta+\theta')}}{[L_{d}(\xi)]^2}, \nonumber \\
     &= \frac{1}{\pi}\int_{0}^{\pi}{\rm d}\xi \frac{2\cos{\xi \theta}\cos{\xi \theta'}}{[L_{d}(\xi)]^2}.
 \end{align}

If $t=t'$, the expression of $\langle \Delta A_{\theta}(t) \Delta A_{\theta'}(t') \rangle$ becomes
\begin{align}
    \label{eq:correlator_all_time_scales}
    & \langle \Delta A_{\theta}(t) \Delta A_{\theta'}(t) \rangle  \\
    & = \frac{2D}{2\pi}   \int_{-\pi}^{\pi}{\rm d}\xi \frac{e^{i\xi(\theta-\theta')} + e^{i\xi(\theta+\theta')}}{[L_{d}(\xi)]^2} \left\{\Delta t + \frac{\tau}{L_{d}(\xi)} \right. \nonumber \\
    & \hspace{5cm} \left.\left( e^{-\frac{L_{d}(\xi)}{\tau}\Delta t}-1 \right) \right\}, \nonumber \\
    &\xrightarrow[\tau \mapsto 0]{} 2D \Delta t \mathcal{D}_2(\theta,\theta'). \nonumber
\end{align}
Note that in the other limit $\tau \gg 1$, one finds that 
\begin{equation} \label{eq:epps}
    \langle \Delta A_{\theta}(t) \Delta A_{\theta'}(t) \rangle \approx D \frac{\Delta t^2}{\tau} \mathcal{D}_1(\theta,\theta')
\end{equation}
This encodes the Epps effect: correlations tend to zero at very small time resolutions.

Otherwise, if $|t-t'|> \Delta t$, we obtain
\begin{align}
    &\langle \Delta A_{\theta}(t) \Delta A_{\theta'}(t') \rangle \nonumber \\
    &= -\frac{\tau D \Delta t}{\pi} \int_{-\pi}^{\pi}{\rm d}\xi \frac{e^{i\xi(\theta-\theta')} + e^{i\xi(\theta+\theta')}}{[L_{d}(\xi)]^3} \hspace{3cm}\nonumber \\
    & \hspace{3cm} \left(e^{-\frac{L_{d}(\xi)}{\tau}(t-t')} +e^{-\frac{L_{d}(\xi)}{\tau}(t-t' -\Delta t)}\right) \nonumber \\
    & \xrightarrow[\tau \mapsto 0]{} 0
\end{align}

Hence, for $\tau$ close to $0$, the covariance of $\Delta A$ can be written as
 \begin{equation}
    \langle \Delta A_{\theta}(t) \Delta A_{\theta'}(t') \rangle =
    \left\lbrace
    \begin{aligned}
    &0, \text{ if } |t-t'|> \Delta t, \\
    &2D \Delta t \mathcal{D}_2(\theta,\theta'), \text{ if } t=t'.
    \end{aligned}
    \right.
\end{equation}

\section{Closed formulas for the correlators when $\psi \gg 1$} \label{Closed formulas for the correlators}
We perform a rotation of the tenors $\theta$ and $\theta'$ in order to formulate our model in relation to the diagonal and anti-diagonal elements of the variance-covariance matrix.
\begin{equation}
    \left\lbrace
    \begin{aligned}
    &\theta_+ = \theta + \theta', \\
    &\theta_- = \theta - \theta',
    \end{aligned}
    \right.
\end{equation}
For integer $\theta$'s, a second change of variable $z=e^{i\xi}$ allows us to write $\mathcal{D}_k$ as a contour integral
\begin{align}
    \mathcal{D}_k(\theta_+;\theta_-) = \frac{\nu^4}{2\pi} \int_{\gamma(1)} \frac{z^{|\theta_-|+1}+z^{\theta_++1}}{P(z)^{k}},
\end{align}
where $P(z)=z^2 - z\left(\frac{z-1}{\mu}\right)^2+\left(\frac{z-1}{\nu}\right)^4$ and $\gamma(1)$ is the unit circle. The residue theorem then yields
\begin{align}
    \mathcal{D}_1(\theta,\theta') = \nu^4 \frac{(\beta^{+}_{-})^{|\theta_-|+1}+(\beta^{+}_{-})^{\theta_++1}}{(\beta^{+}_{-}-\beta_{+}^{+})(\beta^{+}_{-}-\beta^-_+)(\beta^{+}_{-}-\beta_{-}^{-})} \nonumber\\
    + \nu^4 \frac{(\beta_{-}^{-})^{|\theta_-|+1}+(\beta_{-}^{-})^{\theta_++1}}{(\beta_{-}^{-}-\beta_{+}^{+})(\beta_{-}^{-}-\beta_{+}^{-})(\beta_{-}^{-}-\beta_{-}^{+})},
\end{align}
where,
\begin{align}
    \beta^+_{\pm} =1 + \frac{\alpha_+ \pm \sqrt{\alpha_+(\alpha_+ + 4)}}{2}, \nonumber\\
    \beta^-_{\pm} =1 + \frac{\alpha_- \pm \sqrt{\alpha_-(\alpha_- + 4)}}{2}, \nonumber\\
    \alpha_{\pm} = \frac{\nu^4}{2\mu^2} \left(1 \pm \sqrt{1-4\left(\frac{\mu}{\nu}\right)^4} \right).
\end{align}

Similarly, one computes $\mathcal{D}_2(\theta,\theta')$ thanks to the residue theorem:
\begin{align}
    \mathcal{D}_2(\theta,\theta') = \text{Res}(g,\beta^{+}_{-}) + \text{Res}(g,\beta_{-}^{-}),
\end{align}
where $g(z) = \nu^8\frac{z^{|\theta_-|+3}+z^{\theta_++3}}{P(z)^2}$ and the residuals $\text{Res}(g,\beta^{+}_{-})$ and $\text{Res}(g,\beta_{-}^{-})$ at the poles $\beta^{+}_{-}$ and $\beta^{-}_{-}$ are given by
\begin{align}
    &\text{Res}(g,\beta^{+}_{-}) =  \nu^8\frac{(\beta^{+}_{-})^{|\theta_-|+2}+(\beta^{+}_{-})^{\theta_++2}}{(\beta^{+}_{-}-\beta_{+}^{+})^2(\beta^{+}_{-}-\beta^{-}_{+})^2(\beta^{+}_{-}-\beta_{-}^{-})^2} \nonumber \\
    & \hspace{2cm} + \nu^8\frac{(\beta^{+}_{-})^{|\theta_-|+3}+(\beta^{+}_{-})^{\theta_++3} }{(\beta^{+}_{-}-\beta_{+}^{+})^3(\beta^{+}_{-}-\beta^{-}_{+})^3(\beta^{+}_{-}-\beta_{-}^{-})^3} \nonumber \\
    & \hspace{2cm} \left((\beta^{+}_{-}-\beta^{-}_{+})(\beta^{+}_{-}-\beta^{-}_{-}) + (\beta^{+}_{-}-\beta^{+}_{+})(\beta^{+}_{-}-\beta^{-}_{-}) \right. \nonumber \\ 
    & \hspace{2cm} \left.+ (\beta^{+}_{-}-\beta^{+}_{+})(\beta^{+}_{-}-\beta^{-}_{+})  \right),
\end{align}
and by
\begin{align}
    &\text{Res}(g,\beta^{-}_{-}) =  \nu^8\frac{(\beta^{-}_{-})^{|\theta_-|+2}+(\beta^{-}_{-})^{\theta_++2}}{(\beta^{-}_{-}-\beta_{+}^{+})^2(\beta^{-}_{-}-\beta^{-}_{+})^2(\beta^{-}_{-}-\beta^{+}_{-})^2} \nonumber \\
    & \hspace{2cm}  + \nu^8\frac{(\beta^{-}_{-})^{|\theta_-|+3}+(\beta^{-}_{-})^{\theta_++3} }{(\beta^{-}_{-}-\beta_{+}^{+})^3(\beta^{-}_{-}-\beta^{-}_{+})^3(\beta^{-}_{-}-\beta^{+}_{-})^3} \nonumber \\
    & \hspace{2cm}  \left((\beta^{-}_{-}-\beta^{-}_{+})(\beta^{-}_{-}-\beta^{+}_{-}) + (\beta^{-}_{-}-\beta^{+}_{+})(\beta^{-}_{-}-\beta^{+}_{-}) \right. \nonumber \\ 
    & \hspace{2cm}  \left.+ (\beta^{-}_{-}-\beta^{+}_{+})(\beta^{-}_{-}-\beta^{-}_{+})  \right).
\end{align}

\section{Solution to the discretized Master Equation when $\psi \ll 1$} \label{Solution to the discretized Master Equation psi below one}
To ensure the validity of the method of images we need $A_{-\theta}$ to be a solution of the unconstrained problem if $A_{\theta}$ is a solution. Therefore, we assume that $\psi \ll \theta$. As shown below, in this case, the model can be written as a function of the products $\mu\psi$ and $\nu\psi$ only. Hence, one can choose the parameter $\psi$ to be arbitrarily small to ensure that $\psi \ll 1 \leq \theta$. In this limit, the matrix $\mathcal{M}$ in Eq.~\eqref{eq:discret_master_general} becomes:
\begin{align}
    \mathcal{M}_{\theta\theta'} &:= (I)_{\theta\theta'} \nonumber \\
    &+ \theta\left(\frac{1}{\kappa_2^4}- \frac{1}{\kappa_1^2}\right)\left(\frac{1}{2}I_1 - \frac{1}{2}I_{-1} \right)_{\theta\theta'} \nonumber \\
    &+ \theta^2\left(\frac{7}{\kappa_2^4}- \frac{1}{\kappa_1^2}\right) \Big( I_1 - 2I +I_{-1}\Big)_{\theta\theta'}  \nonumber \\
    &+ \frac{6\theta^3}{\kappa_2^4} \left(\frac{1}{2}I_{2} - I_1 + I_{-1} - \frac{1}{2}I_{-2}\right)_{\theta\theta'}  \nonumber \\ 
&+ \frac{\theta^4}{\kappa_2^4} \Big( I_{2} - 4I_1 + 6I -4I_{-1} + I_{-2} \Big)_{\theta\theta'} .
\end{align}
where we note $\kappa_1$ and $\kappa_2$ for  $\mu\psi$ and $\nu\psi$ respectively, without loss of generality. 

For $(\theta,\theta') \in \mathbb{Z}$ we define the \textit{propagator matrix}~$\mathcal{G}_{\theta\theta'}(t,t')$ as the solution to
\begin{align}
    \label{eq:numerical_propagator}
    \frac{\partial \mathcal{G}}{\partial t}(t-t') + \frac{1}{\tau} \mathcal{M} \mathcal{G}(t-t') = \delta(t-t') I.
\end{align}
The propagator reads
\begin{align}
 \mathcal{G}(t-t') = H(t-t') e^{-\frac{t-t'}{\tau}\mathcal{M}}.
\end{align}
Let $n$ be the number of available SOFR futures. We assume that the matrix $\mathcal{M}$ is of size $2N+1$ with $N \gg n$. We define the indices of the lines and columns of $\mathcal{M}$ in the range $[-N,N]$. 

The matrix $\mathcal{M}$ is invariant by a central symmetry with respect to the center of the matrix $\mathcal{M}_{00}$:
\begin{align}
    \mathcal{M} = \Bar{I} \mathcal{M} \Bar{I},
\end{align}
where $\bar{I}$ is a matrix with ones only on its largest anti-diagonal. Hence, $\mathcal{F}_{\theta\theta'}(t-t') := \mathcal{G}_{-\theta\theta'}(t-t')$ is a solution for $(\theta,\theta') \in \mathbb{Z}$ of the following equation: 
\begin{align}
    \frac{\partial \mathcal{F}}{\partial t}(t-t')  + \frac{1}{\tau} \mathcal{M}\mathcal{F}(t-t') = \delta(t-t')  \bar{I}.
\end{align}
It yields a solution for $(\theta,\theta') \in \mathbb{Z}$:
\begin{align}
 \mathcal{F}(t-t') =  H(t-t') e^{-\frac{t-t'}{\tau}\mathcal{M}} \bar{I}.
\end{align}
By the method of images, the propagator $G$ for $(\theta,\theta') \in \mathbb{Z}^2$ associated to the solution that satisfies the Neumann boundary condition, is given by:
\begin{align}
    G(t-t')&:= H(t-t')  e^{-\frac{t-t'}{\tau}\mathcal{M}} \left( I + \bar{I} \right).
\end{align}
As the bottom left (i.e. $\theta > 0$ and $\theta' < 0$) and top right bloc matrices (i.e. $\theta < 0$ and $\theta' > 0$) of $\mathcal{M}$ are nil, the restriction of the matrix $e^{-\frac{t-t'}{\tau}\mathcal{M}}(I+\bar{I})$ to positive $\theta$ is equal to the product of the restricted matrices $e^{-\frac{t-t'}{\tau}\mathcal{M}}$ and $I+\bar{I}$. We note $\mathcal{J}$ the restriction to positive values of $\theta$ of the matrix $I+\bar{I}$:
\begin{equation}
    \mathcal{J}_{\theta\theta'} =
    \left\lbrace
    \begin{aligned}
    &2, \text{ if } \theta=\theta'=0, \\
    &1, \text{ if } \theta=\theta'>0, \\
    &0, \text{ if } \theta\neq\theta'.
    \end{aligned}
    \right.
\end{equation}

Finally, for $\theta \in \mathbb{N}$, the solution $A$ to Eq.~\eqref{eq:discret_master_general} is given by
\begin{align}
    A(t) &= \frac{1}{\tau}\int_{-\infty}^{t} {\rm d}t' G(t-t') \eta(t') \nonumber \\
    &= \frac{1}{\tau}\int_{-\infty}^{t} {\rm d}t' e^{-\frac{t-t'}{\tau}\mathcal{M}} \mathcal{J} \eta(t')
\end{align}

\section{Noise correlators when $\psi \ll 1$} \label{Noise correlators psi below one}
\subsection{Autocovariance of the correlated noise for $\tau=0$}
For $\tau \to 0$, $e^{-\frac{t-t'}{\tau}\mathcal{M}} \to \tau \mathcal{M}^{-1}\delta(t-t')$. Hence we can derive a closed formula for the noise $A$:
\begin{align}
    A(t) = \mathcal{M}^{-1}\mathcal{J}\eta(t)
\end{align}
Thus, the autocovariance of A is given by 
\begin{align}
    \langle A(t) A(t')^\top\rangle = 2 D \delta(t-t') \mathcal{M}^{-1}\mathcal{J}^2(\mathcal{M}^{-1})^\top.
\end{align}
In this limit, the autocovariance of $\Delta A$, the coarse-grain cumulative sum of $A(t)$ over the time interval $\Delta t \gg \tau$, is given by
\begin{equation}
    \langle \Delta A(t) \Delta A(t')^\top\rangle =
    \left\lbrace
    \begin{aligned}
    &0, \text{ if } |t-t'|> \Delta t, \\
    &2 D \Delta t \mathcal{M}^{-1}\mathcal{J}^2(\mathcal{M}^{-1})^\top, \text{ if } t=t'.
    \end{aligned}
    \right.
\end{equation}
Finally, the equal-time Pearson correlation coefficient among coarse-grained forward rate variations $\Delta f_\theta$ is given by
\begin{align}
    \label{eq:micro-founded_final}
    \rho_{\theta\theta'} = \frac{\left(\mathcal{M}^{-1}\mathcal{J}^2(\mathcal{M}^{-1})^\top\right)_{\theta\theta'}}{\sqrt{\left(\mathcal{M}^{-1}\mathcal{J}^2(\mathcal{M}^{-1})^\top\right)_{\theta\theta}\left(\mathcal{M}^{-1}\mathcal{J}^2(\mathcal{M}^{-1})^\top\right)_{\theta'\theta'}}}.
\end{align}

\subsection{Autocovariance of the correlated noise for $\tau >0$}

We assume $\mathcal{M}$ is diagonalizable. We define $P$ as the transformation matrix of $\mathcal{M}$, i.e.
\begin{align}
    \mathcal{M} := P \Lambda P^{-1} := \sum_{k\in \mathbb{N}} \lambda_k U_k V_k^\top
\end{align}
where the $U_k$ are the column vectors of the matrix $P$, $V_k^\top$ are line vectors of the matrix $P^{-1}$, and $\lambda_k$ are the eigenvalues of $\mathcal{M}$. The noise field $A$ can be expressed in this new basis as
\begin{align}
    A(t) =  \frac{1}{\tau} \sum_{k\in \mathbb{N}} \int_{-\infty}^{t} {\rm d}t' e^{-\frac{t-t'}{\tau}\lambda_k} \eta(t') U_k V_k^\top \mathcal{J}
\end{align}
The autocovariance of $A$ reads
\begin{align}
    &\langle A(t)A(t')^\top \rangle  \nonumber \\ 
    & = \frac{2D}{\tau^2}\sum_{(k,k')\in \mathbb{N}^2} U_k V_k^\top \mathcal{J}^2 V_{k'} U_{k'}^\top \nonumber \\  
    & \hspace{3cm}\int_{-\infty}^{\min(t,t')} {\rm d}u  e^{-\frac{t-u}{\tau}\lambda_k-\frac{t'-u}{\tau}\lambda_{k'}}, \nonumber \\
    & = \frac{2D}{\tau}\sum_{(k,k')\in \mathbb{N}^2}  \frac{K_{kk'}}{\lambda_k+\lambda_{k'}}e^{\frac{1}{\tau}\left(-\lambda_k t-\lambda_{k'}t' +(\lambda_k+\lambda_{k'})\min{(t,t')}\right)},
\end{align}
where the matrices $K_{kk'}$ are defined by $K_{kk'}:= U_k V_k^\top \mathcal{J}^2 V_{k'} U_{k'}^\top$. We now recall the definition of the cumulative sum of $A$ over an interval of lenght $\Delta t$:
\begin{align}
    \Delta A(t) := \int_{t-\Delta t}^t {\rm d}u A(u)
\end{align}
Hence, the autocovariance of $\Delta A$ reads
\begin{align}
    & \langle \Delta A(t) \Delta A(t')^\top \rangle \nonumber \\
    &= \int_{t-\Delta t}^t {\rm d}u\int_{t'-\Delta t}^{t'}   {\rm d}u' \langle A(u)A(u')^\top \rangle \nonumber \\
    &= \frac{2D}{\tau}\sum_{(k,k')\in \mathbb{N}^2}  \frac{K_{kk'}}{\lambda_k+\lambda_{k'}} \nonumber \\
    & \hspace{1cm}\int_{t-\Delta t}^t {\rm d}u\int_{t'-\Delta t}^{t'}   {\rm d}u' e^{\frac{1}{\tau}\left(-\lambda_k u-\lambda_{k'}u' +(\lambda_k+\lambda_{k'})\min{(u,u')}\right)}
\end{align}
For $t=t'$, we have, 
\begin{align}
    \label{eq:correlator_all_time_scales_psi_below_one}
    &\langle \Delta A(t) \Delta A(t)^\top \rangle = 2D \sum_{(k,k')\in\mathbb{N}^2} \frac{K_{kk'}}{\lambda_k+\lambda_{k'}} \Bigg\{ \nonumber \\  
    & \Delta t (\frac{1}{\lambda_k} + \frac{1}{\lambda_{k'}}) + \tau\left(\frac{1}{\lambda_k^2}(e^{-\frac{\Delta t}{\tau}\lambda_k}-1) + \frac{1}{\lambda_{k'}^2}(e^{-\frac{\Delta t}{\tau}\lambda_{k'}}-1) \right) \Bigg\}.
\end{align}
Notably, in the stationary limit, for $\tau$ close to zero,
\begin{align}
    \langle \Delta A(t) \Delta A(t)^\top \rangle &=2D \sum_{(k,k')\in\mathbb{N}^2}K_{kk'} \frac{1}{\lambda_k\lambda_{k'}} \Delta t
\end{align}
In the other limit $\tau \gg 1$, one finds that 
\begin{equation} \label{eq:epps_numerical}
    \langle \Delta A_{\theta}(t) \Delta A_{\theta'}(t) \rangle \approx 2D \sum_{(k,k')\in\mathbb{N}^2}K_{kk'} \frac{\Delta t^2}{\tau} \frac{\lambda_k^2 + \lambda_{k'}^2}{(\lambda_k + \lambda_{k'})\lambda_k^2\lambda_{k'}^2}
\end{equation}
This encodes the Epps effect: correlations tend to zero at very small time resolutions.

\section{The Baaquie-Bouchaud, Logarithm}
The stiff propagator BB04 model \citep{BaaquieBouchaud-2004} proposed the change of variable $\bar{z}(\theta) = \theta^{\bar{\psi}}$. As mentioned in section~\ref{Psychological_time}, this formulation violates the constraint that for very small maturities psychological time and real time should become equivalent. In addition, this change of variable is mis-specified for $\psi$ close to zero. Indeed, for fixed $\theta$
\begin{align}
     \bar{z}(\theta) \xrightarrow[\psi\mapsto 0]{} 1.
\end{align}
Remediating these two limitations requires introducing two parameters, $\psi$ and $\zeta$, in the definition of the psychological time:
\begin{align}
    \bar{z}(\theta) = \frac{\psi}{\zeta}\left( \left(1+\frac{\theta}{\psi}\right)^{\zeta} -1\right),
\end{align}
where ${\psi}$ has dimensions of time and $\zeta$ is a pure number $\leq 1$.
For this new change of variable $\bar{z}(\theta)$ is equivalent to $\theta$ for $\theta$ approaching $0$, proportional to $\theta^{\zeta}$ for large values of $\theta$, and equal to $\theta$ for $\zeta=1$. Moreover, for $\zeta$ approaching $0$,
\begin{align}
    \bar{z}(\theta)  \approx \psi \log{\left(1+\frac{\theta}{\psi} \right)}  = z(\theta)
\end{align}
Actually, we found that the calibration of the BB04 with the change of variable $\bar{z}$ yields an optimal value for $\zeta$ very close to $0$. Hence we define a regularized version of the BB04 model by replacing $\bar{z}(\theta)$ by $z(\theta)$. For such a specification, the equal-time Pearson correlation is given by
\begin{align}
    \label{eq:BBL}
    \Tilde{\rho}_{\theta\theta'} = \frac{\mathcal{D}_{BB}(z(\theta),z(\theta'))}{\sqrt{\mathcal{D}_{BB}(z(\theta),z(\theta))\mathcal{D}_{BB}(z(\theta'),z(\theta'))}},
\end{align}

a result we will refer to as BBL3 model for Baaquie-Bouchaud, Logarithm, three parameters.

\section{Two-parameter versions} \label{Two-parameters versions}
In this section, we compare the performances of the two-parameter variants of our models by assigning to the stiffness parameter $\nu$ an infinite value. This adjustment applies to: (i) the regularized version BBL3 of the continuous model from \cite{BaaquieBouchaud-2004}, using Eq.~\eqref{eq:BBL}; (ii) our micro-founded discrete model BBD3, using Eq.~\eqref{eq:noise_correlator_BBD3}. These models are denoted as BBL2 and BBD2 respectively.

Optimal calibration parameters, obtained by fitting these models to empirical Pearson correlations for the period~$1994-2023$, are displayed in Table~\ref{tab:fitted_param2}. The BBD2 model presents plausible values for the psychological time and line tension parameters. In contrast, the BBL2 model results in improbable values for $\psi$ and $\mu$. As previously mentioned, this is primarily because the correlation surface develops a cusp around the diagonal $\theta=\theta'$, which was actually the very reason why \citet{BaaquieBouchaud-2004} introduced the stiffness term $\nu$.

\begingroup
\squeezetable
\begin{table}[h]
    \begin{ruledtabular}
    \begin{tabular}{ccccc}
    Model & $\psi^*$ (months) & $\mu^*$\\
    \hline
    BBL2 & $1.27\times 10^{-5}$ & $5.21 \times 10^{4}$  \\
    BBD2 & $2.00$ & $1.01$  \\
    \end{tabular}
    \end{ruledtabular}
    \caption{Optimal parameters obtained when fitting the tested models to empirical Pearson correlations for the period~$1994-2023$. Two models are considered: (i) the continuous regularized model BBL2 (Eq.~\eqref{eq:BBL} with $\nu=\infty$) and our micro-founded discrete model BBD2 (Eq.~\eqref{eq:noise_correlator_BBD3} with $\nu=\infty$). While $\mu$ is dimensionless in the discrete models, this parameter is in units of $\text{3 months}^{-1}$ in the case of the BBL2 model. \label{tab:fitted_param2}}
\end{table}
\endgroup

Figure~\ref{fig:argest_anti_diag_['BBL2', 'BBD2']_full_period} depicts the correlation coefficients along the most extended anti-diagonal for the period~$1994-2023$ as determined by the calibration of the BBL2 and BBD2. It also illustrates the typical error across the correlation surface, underscoring the superior precision of the BBD2 model relative to the continuous variant. 
\begin{figure}
    \centering
    \includegraphics[width=0.95\linewidth]{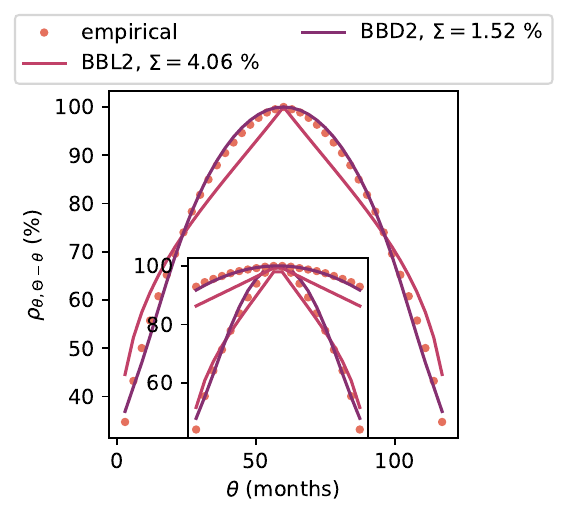}
    \caption{Dots represent the empirical correlation $\rho_{\theta\theta'}$ along the longest stretch perpendicular to the diagonal, i.e. $\theta' = \Theta - \theta$ , where $\Theta$ is the maximum available maturity. The plain lines are the best fit for: (i) the regularized version BBL2 of the continuous model \cite{BaaquieBouchaud-2004}, using Eq.~\eqref{eq:BBL} with $\nu=\infty$; our micro-founded discrete model BBD2, using Eq.~\eqref{eq:noise_correlator_BBD3} with $\nu=\infty$.}
    \label{fig:argest_anti_diag_['BBL2', 'BBD2']_full_period}
\end{figure}

\section{Epps effect when $\psi \gg 1$} \label{Epps effect psi below one}

\begin{figure}
    \centering
    \includegraphics{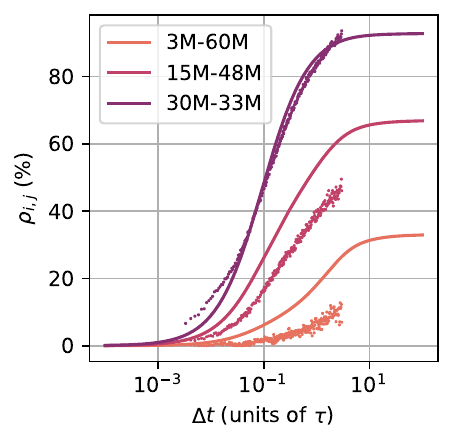}
    \caption{Plain lines: theoretical Pearson correlation coefficients among three pairs of forward rate variations~$(\Delta f_\theta, \Delta f_{\theta'})$ as a function of the time scale~$\Delta t$ (see~\eqref{eq:correlator_all_time_scales}).  Using the empirical correlations of the pair $30$-$33$ months, the  parameter~$\varepsilon$ of the idiosyncratic white noise was calibrated to $0.026$, and the characteristic time of the Epps effect$\tau$ to $21$ minutes. This figure also shows the theoretical correlations yielded by this set of parameters for two other pairs ($15$-$48$ and $3$-$60$ months).
    Dots: empirical Pearson correlation coefficients for three pairs of SOFR Futures prices for the year $2021$ at time scales ranging from $4$ seconds to one hour.}
    \label{fig:epps_theo_imf_2_log}
\end{figure}

Each colored line in Fig.~\ref{fig:epps_theo_imf_2_log} represents the correlation $\rho_{\theta \theta'}$ across different time scales~$\Delta t$ among pairs of forward rate variations $(\Delta f_\theta, \Delta f_{\theta'})$, as given by our model in the case $\psi \gg 1$ (see Eq.~\eqref{eq:correlator_all_time_scales}) calibrated on daily correlations (see section~\ref{Calibration over the whole sample}) with the additional fitting parameter~$ \varepsilon$. For the pair $30$-$33$  months, we find $\varepsilon \approx 0.026$ ($\mathcal{C}(\theta,\theta')$ is in the range of $0.14$ to $0.48$ when $\psi \gg 1$) and $\tau \approx 21$ minutes. Fig.~\ref{fig:epps_theo_imf_2_log} demonstrates that our model is able to reproduce the whole dependence of the empirical correlations of pairs of SOFR Futures binned at different time scales.

\section{Curvature along the anti-diagonals} 
\label{Curvature along the anti-diagonal}
One of the most salient successes of the BB04 model is its ability, in line with observations, to reproduce the power-law decay of the curvature of forward rate correlations perpendicular to the diagonal. Fig.~\ref{fig:curvature_power_law_full_period_['BBL3', 'BBD3', 'BBD2']} shows estimations of the curvatures generated by the tested models and the ones empirically observed. These estimations are produced through the fitting of parabolas using $10$ points around the center of each anti-diagonal of the correlation surface for the $1994-2023$ period. Fig.~\ref{fig:curvature_power_law_full_period_['BBL3', 'BBD3', 'BBD2']} reveals the adequacy of the continuous model BBL3, and the discrete models, BBD3, BBD2, and BBDL, with the observed curvature. However, one can notice a slight change of convexity in the curvature for BBD3 and BBD2: this is probably due to the change of variable in $z(\theta)$ in Eq.~\ref{eq:noise_correlator_BBD3}.
\begin{figure}
    \centering
    \includegraphics[width=\linewidth]{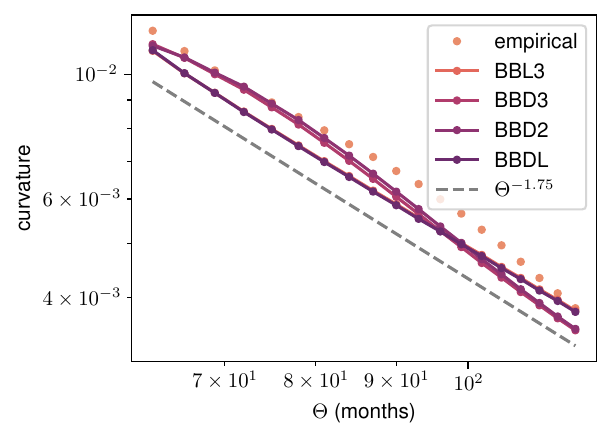}
    \caption{Curvature of the correlation surface along the stretches perpendicular to the diagonal, i.e. $\theta' = \Theta - \theta$ as a function of largest tenor $\Theta$ for the $1994-2023$ period. Here we compare the continuous regularized model BBL3 (Eq.~\eqref{eq:BBL}) with our three micro-founded discrete models: BBD3 (Eq.~\eqref{eq:noise_correlator_BBD3}), BBD2 (Eq.~\eqref{eq:noise_correlator_BBD3} with $\nu \to \infty$), and BBDL (Eq.~\eqref{eq:noise_correlator_BBDL}).} 
    \label{fig:curvature_power_law_full_period_['BBL3', 'BBD3', 'BBD2']}
\end{figure}

\end{document}